\documentclass[letterpaper]{article} % DO NOT CHANGE THIS
\usepackage{aaai25}  % DO NOT CHANGE THIS % DO NOT CHANGE THIS
\usepackage{times}  % DO NOT CHANGE THIS
\usepackage{helvet}  % DO NOT CHANGE THIS
\usepackage{courier}  % DO NOT CHANGE THIS
\usepackage[hyphens]{url}  % DO NOT CHANGE THIS
\usepackage{graphicx} % DO NOT CHANGE THIS
\usepackage{natbib}  % DO NOT CHANGE THIS AND DO NOT ADD ANY OPTIONS TO IT
\usepackage{caption} % DO NOT CHANGE THIS AND DO NOT ADD ANY OPTIONS TO IT
\usepackage{times}
\usepackage{soul}
\usepackage{url}
\usepackage[utf8]{inputenc}
\usepackage{graphicx}
\usepackage{amsmath}
\usepackage{amsthm}
\usepackage{booktabs}
\usepackage{algorithm}
\usepackage{algorithmic}
\usepackage[switch]{lineno}

\usepackage[utf8]{inputenc} % allow utf-8 input
\usepackage[T1]{fontenc}    % use 8-bit T1 fonts
\usepackage{url}            % simple URL typesetting
\usepackage{amsfonts}       % blackboard math symbols
\usepackage{nicefrac}       % compact symbols for 1/2, etc.
\usepackage{microtype}      % microtypography
\usepackage{fancyhdr}       % header

\usepackage{xspace}
\usepackage{bm}
\usepackage{siunitx}

\usepackage{multirow}
\usepackage{float}

\usepackage{makecell}
\usepackage{cleveref}
\usepackage[caption=false]{subfig}

\usepackage{mathtools} 
\usepackage{xcolor}
\usepackage{amsthm}

\usepackage[utf8]{inputenc} % allow utf-8 input
\usepackage[T1]{fontenc}    % use 8-bit T1 fonts
\usepackage{url}            % simple URL typesetting
\usepackage{amsfonts}       % blackboard math symbols
\usepackage{nicefrac}       % compact symbols for 1/2, etc.
\usepackage{microtype}      % microtypography
\usepackage{fancyhdr}       % header

\usepackage{xspace}
\usepackage{bm}
\usepackage{siunitx}

\usepackage{multirow}
\usepackage{float}

\usepackage{makecell}
\usepackage{cleveref}
\usepackage[caption=false]{subfig}

\usepackage{mathtools} 
\usepackage{xcolor}
\usepackage{amsthm}

\newcolumntype{P}[1]{>{\centering\arraybackslash}p{#1}}

\DeclareCaptionStyle{ruled}{labelfont=normalfont,labelsep=colon,strut=off} % DO NOT CHANGE THIS
\title{Who Pays the RENT? Implications of Spatial Inequality for Prediction-Based Allocation Policies}

\author {
    Tasfia Mashiat\textsuperscript{\rm 1,}\footnote{Work was performed while the author was a PhD student at George Mason University.},
    Patrick J. Fowler\textsuperscript{\rm 2},
    Sanmay Das\textsuperscript{\rm 3}
}
\affiliations {
    \textsuperscript{\rm 1}University of Iowa\\
    \textsuperscript{\rm 2}Washington University in St. Louis\\
    \textsuperscript{\rm 3}Virginia Tech\\
    
    tasfia-mashiat@uiowa.edu, pjfowler@wustl.edu, sanmay@vt.edu
}

\begin{document}
\maketitle

\begin{abstract} 
AI-powered scarce resource allocation policies rely on predictions to target either specific individuals (e.g., high-risk) or settings (e.g., neighborhoods). Recent research on individual-level targeting demonstrates conflicting results; some models show that targeting is not useful when inequality is high, while other work demonstrates potential benefits. 
To study and reconcile this apparent discrepancy, we develop a stylized framework based on the Mallows model to understand how the spatial distribution of inequality affects the effectiveness of door-to-door outreach policies. We introduce the RENT (Relative Efficiency of Non-Targeting) metric, which we use to assess the effectiveness of targeting approaches compared with neighborhood-based approaches in preventing tenant eviction when high-risk households are more versus less spatially concentrated.
We then calibrate the model parameters to eviction court records collected in a medium-sized city in the USA. Results demonstrate considerable gains in the number of high-risk households canvassed through individually targeted policies, even in a highly segregated metro area with concentrated risks of eviction. 
We conclude that apparent discrepancies in the prior literature can be reconciled by considering 1) the source of deployment costs and 2) the observed versus modeled concentrations of risk. Our results inform the deployment of AI-based solutions in social service provision that account for particular applications and geographies.

\end{abstract}

% %%%%%%%%%%%%%%%%%%%%%%%%%%%%%%%%%%%%%%%%%%%%%%%%%%%%%%%%%%%%%%%%%%%%%%%%

% %%% Include any author-defined commands here.
         
% \newcommand{\BibTeX}{\rm B\kern-.05em{\sc i\kern-.025em b}\kern-.08em\TeX}

% %%%%%%%%%%%%%%%%%%%%%%%%%%%%%%%%%%%%%%%%%%%%%%%%%%%%%%%%%%%%%%%%%%%%%%%%

%%% The following commands remove the headers in your paper. For final 
%%% papers, these will be inserted during the pagination process.

% \pagestyle{fancy}
% \fancyhead{}

%%% The next command prints the information defined in the preamble.

\maketitle 

%%%%%%%%%%%%%%%%%%%%%%%%%%%%%%%%%%%%%%%%%%%%%%%%%%%%%%%%%%%%%%%%%%%%%%%%

\section{Introduction}

Prediction-based resource allocation systems calibrate available task-relevant information to make targeted decisions for individuals in different societal service provisions, such as homelessness, education, medical care, child welfare, etc.~\cite{chouldechova2018case,kube2019allocating,bruce2011track,ayer2019prioritizing}. Generally, the prediction produces a score for each individual that indicates the need for intervention. Scoring informs decision-making in the context of perceived needs among the overall population. %Such scoring allows decision-makers to make informed decisions knowing the distribution of responses received by the overall population. 
In limited resource allocation settings, scores provide a rank for targeting interventions until the budget runs out. %are used to rank individuals to target intervention until the budget runs out. 

Alternatively, communities often rely on non-targeting allocation methods that proportion resource provision based on the aggregated welfare of a larger unit containing multiple individuals.
A motivating example for this paper involves door-to-door canvassing in neighborhoods with high rates of tenant eviction. Canvassers approach all residents in an area to offer information on available resources, regardless of housing precarity or the need for services. Such outreach strategies are easier and faster to coordinate than prediction-based allocations~\cite{moon2024human,johnson2022bureaucratic}; however, for relatively low base-rate events, like eviction filing, the lack of targeting proves inefficient and potentially wasteful. 
 
Although the selection process of units of targeted versus non-targeted allocation systems differs, the underlying goal remains the same -- \textit{Within a given budget, serve as many needy individuals as possible}. 

Recent work of \citet{shirali2024allocation} raises interesting questions regarding the relative utility of prediction-based versus non-targeting allocation methods. In their proposed mathematical model, they 
show that prediction-based targeting policies outperform non-targeting policies for allocating interventions only when the across-unit inequality of welfare is low and the allocation budget is high. Otherwise, non-targeted allocation outperforms prediction-based individualized allocation in all other social welfare and budget contexts.

We note two key model assumptions that introduce potential alternatives that limit generalizations made by Shirali et al. First, the authors consider two categories of allocation costs in their conceptualization: the cost-to-target and the cost-to-treat for each individual. In their work, the modeling of non-targeting policies only includes cost-to-treatment, while targeted policies incorporate both costs. Thus, the model cannot easily capture the costs of service delivery to different locations in situations where the inequality is spatially heterogeneous. Second, Shirali et al. assume that resource allocators possess prior knowledge regarding which individuals exhibit the highest levels of welfare and refrain from allocating interventions to such individuals, even under conditions of non-targeted allocation. Together, the assumptions impose a form of targeting upon specific individuals, despite the intention of non-targeted allocation across units. The assumptions fit certain resource allocation scenarios, for example, allocation of resources to schools in order to treat specific students who may need additional assistance \cite{perdomo2023difficult}. However, they miss essential features of other scenarios, such as door-to-door outreach, where the marginal cost of treating an additional unit depends on spatial location relative to other at-risk units~\cite{mashiat2024beyond}.

In this paper, we focus on the door-to-door outreach setting. Individuals (or households, we use the terms interchangeably in this paper) reside within specific spatial locations, and costs are associated with traveling from one household to another. As described, our motivating example is the allocation of outreach case workers to canvass residential properties located in neighborhoods to prevent tenant eviction~\cite{mashiat2024beyond}. Each of these properties has an eviction risk score and a geospatial location. We assess the success of allocation policies based on the distribution of High-Risk properties across neighborhoods. We introduce a novel metric, RENT (Relative Efficiency of Non-Targeting), and a stylized framework based on the classic Mallows model of ranking to evaluate the effectiveness of outreach allocation policies for preventing evictions. 

Our results demonstrate the connection between the spatial distribution of risk and the overall efficiency of targeting. In particular, accounting for the cost of outreach,
% \textcolor{red}{without prior knowledge?}, 
targeting policies substantially outperform non-targeting policies in reaching more at-risk households when inequality is spatially dispersed. This raises the question of what level of dispersion is necessary to demonstrate the effectiveness of targeting policies.
% \textcolor{red}{for what? outperformance?}. 
We introduce a method for calibrating our stylized framework to real-world data and analyze eviction records from St. Louis (one of the most segregated cities in the United States on both racial and socioeconomic measures)~\cite{cooperman2014story}. Surprisingly, we find that the Mallows model dispersion parameters corresponding to the data from St. Louis are quite high, indicating that even when segregation is high by standard measures (and thus, inequality of risk is concentrated by a lay definition), targeting adds value for achieving the best use of limited social service resources, canvassing a higher number of risky properties thus reducing evictions than non-targeted measures. 

\subsection{Our Contributions}
\begin{itemize}
    \item We introduce a stylized framework based on the classic Mallows model of ranking for generating spatial distributions of inequality with different concentrations of risk across neighborhoods. 

    \item We use the framework to study the relative effectiveness of non-targeting methods versus prediction-based targeting methods, which we call the RENT metric. 

    \item Leveraging a community partnership, we
    calibrate the stylized model and evaluate the performance of targeting and non-targeting policies in St. Louis, MO.  
    Surprisingly, even in a highly segregated city, the dispersion parameter that best models the risk distribution is high, showing that targeting still has significant value and utility. 
    % \textcolor{red}{redundant with paragraph above?}
     
    % \item 
\end{itemize}

% \subsection{Our Results}
\section{Literature Review}
\subsection{Risk-based Intervention Allocation}
Individual targeting-based allocation policies are widely applied in various resource allocation domains. Targeting policies often rely on predictive models that derive risk scores based on how well individuals would do contingent on receiving the intervention~\cite{kube2019allocating}, or on balancing the expected favorable outcome of an intervention against its associated cost in dynamic settings~\cite{tang2023learning}.
%propose an online allocation framework in which both interventions and individuals arrive sequentially over time. Their dual-price policy - which balances  - matches individuals to the most suitable interventions through targeted allocation.  } 

However, targeting can also be based on the vulnerability of individuals receiving the intervention; the risk scores indicate vulnerability~\cite{pokharel2024discretionary,elster1992local}. The vulnerability scores are then used to form a ranking or assign a category (e.g., high-risk, low-risk). Based on a preference metric/ prioritization scheme set by intervention providers, individuals are selected to receive the intervention. Another conceptualization, which is particularly relevant in tenant eviction, considers risk as the probability of experiencing a bad event (e.g., eviction, emergency room visits, bankruptcy, etc.) in a future time frame~\cite{ayer2019prioritizing,deo2013improving,bastani2021efficient,lee2019optimal}.

In the operations research community, one popular paradigm is ``predict-then-optimize'', referring to first utilizing a predictive model and ensuring its accuracy, then optimizing the outcome or the allocation within a budget~\cite{vanderschueren2022predict}. In allocating homeless services, \citet{kube2019allocating} propose a similar approach in which they allocate counterfactual services to each household and identify the service for which the probability of returning to homelessness is the lowest. Work on decision-focused learning adds the allocation decisions into modeling prediction, such that the predictive model is aware of the need for allocation~\cite{ren2024decision,cortes2024decision,elmachtoub2020decision,shah2022decision,mukhopadhyay2017prioritized,wilder2019melding}.

In a different setting, a risk prediction tool, the Dropout Early Warning System (DEWS)~\cite{bruce2011track}, ranks students according to their school dropout likelihood score. Based on this ranking, the system is designed to choose a threshold and categorize students with a probability of graduation below the threshold as needing interventions, compared to those with scores above the thresdhold~\cite{perdomo2023difficult}. However, Perdomo and colleagues show that modeling the dropout risk individually does not improve the outcome compared to targeting a broader scale. Interestingly, \citet{mashiat2024beyond} find that a risk-based outreach routing policy outperforms several baselines that target tenant eviction prevention at a broader geographic scale.

\subsection{Ranking Generation}

Ranking based on measured scores denotes the relative order of individuals and is often used as a prioritization approach for treatment allocation. Several methods have been proposed to aggregate ranking information, such as the Bradley-Terry, Plackett-Luce, and Mallows models~\cite{bradley1952rank,plackett1975analysis,mallows1957non}. Of these, the Mallows model has been applied most in ranking contexts ~\cite{vitelli2018probabilistic}.

The Mallows model~\cite{mallows1957non} is a fundamental and widely applied approach for the generation and interpretation of ranking data. It is frequently compared to the normal distribution, but is limited to permutations. The model has two key elements: the central order, which represents a chosen ranking of available options, and the dispersion parameter $\phi \in [0, 1]$. The value of $\phi$ influences whether the rankings are more closely concentrated around the central order (i.e., $\phi$ closer to 0) versus more uniformly dispersed across all possible permutations (i.e., $\phi$ closer to 1). 

The Mallows model is popular for generating realistic synthetic data by controlling the parameters. Due to its flexibility in producing arbitrarily large synthetic ranking data, the model can be applied to several problem settings, such as voting~\cite{betzler2014theoretical}, clustering~\cite{busse2007cluster}, matching~\cite{irurozki2019mallows}, and analyzing distributions of ranking data~\cite{lu2014effective,collas2021concentric}. 

In our work, we apply the Mallows model to explore the spatial distribution of rankings in different groups of interest. The group membership of individuals is a representation of settings where individuals are part of a larger unit. For example, students in a school or schools in a school district. We focus on tenant eviction, where residential properties are located in different neighborhoods.

Tenant eviction is a major crisis in the US and has serious negative implications, especially for low-income households. Evicted households experience economic fallout and emotional stress, including
depression, anxiety, and substance abuse~\cite{himmelstein2021eviction,acharya2022risk,bushman2022housing,kim2021financial}. Although governmental and non-governmental agencies coordinate community responses to eviction, efforts often struggle to identify and assist households at risk of eviction with limited and poorly targeted resources~\cite{boen2023buffering,donnelly2021state,leifheit2021variation}. 

Despite the current limitations, the machine learning community is engaging in efforts to address critical challenges. Initiatives include forecasting future evictions by employing machine learning and deep learning frameworks to direct outreach case workers to residential properties that are at high risk of eviction~\cite{tabar2022forecasting,sarkar2024geospatial,mashiat2024beyond}.
However, the effectiveness of these prediction-based targeting policies as a function of the distribution of inequality remains an open question. This problem motivates our work.

\subsection{Spatial Inequality in Resource Allocation}

In societal resource allocation settings, inequality is defined in terms of a discrepancy in the allocation of social goods, such as homeless services, child welfare, and aggregated social welfare~\cite{kube2022just,azizi2018designing,saxena2024algorithmic,shirali2024allocation} among different populations or groups. 
% \textcolor{red}{is this sentence needed?} 
In this work, we focus on spatial inequality, precisely in the context of eviction prevention.

Previous research has reported that factors such as economic disparities, political power dynamics, systemic discrimination, geographic location, etc., can drive spatial inequality~\cite{gligoric2023revealed}. Such inequalities are observed in access to proper healthcare~\cite{mccrum2022disparities}, transportation~\cite{sanchez2003moving}, education~\cite{zahl2024spatial}, etc. For example, reports show discrepancies in the allocation of COVID-19 vaccines between states in the USA based on the existence of a vulnerable population ~\cite{bilal2022heterogeneity}. We also derive a similar notion of spatial inequality between neighborhoods in terms of the distribution of eviction risk and analyze the implications on the effectiveness of outreach. 

% \textcolor{red}{say more?}

\subsection{Reconciling Existing Results}
The two most directly related papers to this one are that of Shirali et al., discussed at length above, and the work of \citet{mashiat2024beyond} on using risk scores to inform eviction outreach. Our work is motivated by the setting of Mashiat et al., in that we model the specific problem of door-to-door outreach for eviction prevention. At the same time, it directly interrogates the key question raised by Shirali et al. -- how does the distribution of inequality (in our case, the inequality is manifested as eviction risk) across space affect the value of targeting? Mashiat et al. report significant value in targeting, while Shirali et al. describe a model where targeting has limited value when inequality is high. We attempt to reconcile these two results and uncover two major factors in the apparent discrepancy. First, the cost of outreach in eviction prevention is such that the marginal cost of outreach is deeply dependent on the spatial distribution of risk, a feature that is not present in the model of Shirali et al. Second, what we might intuitively consider a high inequality regime for eviction risk in practice (the highly segregated neighborhoods of St. Louis) may not correspond with what would mathematically be high inequality in a formal model like the one we present here, based on the Mallows model.

\section{A General Framework}

We now introduce our stylized framework for generating spatial distributions of risk. We use this framework specifically to analyze eviction outreach policies, but it could also be adapted for other types of resource allocation tasks. 

\subsection{Spatial Distribution of Risk or Inequality}
Our model assumes that there are $N$ distinct neighborhoods, each with a fixed number ($n$) of properties (the restriction to each neighborhood having the same number of properties is not necessary or limiting, but simplifies the exposition). 
Each property has a specific \emph{risk score}, and ordering by risk scores yields a ranking of all properties. A cutoff based on the percentile or risk score can be used to differentiate between \emph{High-Risk} and \emph{Low-Risk} properties.

First, consider the case of complete concentration of inequality by neighborhood. In this case, Neighborhood 1 (numbering by risk, without loss of generality) has the $n$ riskiest properties, Neighborhood 2 the next $n$ riskiest, etc. Let us call this the \emph{homogeneous risk ranking}, $\sigma_0$. In order to vary how inequality/risk is distributed across neighborhoods, we use the Mallows model~\cite{mallows1957non}, which takes two parameters, a central ranking 
$\sigma$ and a dispersion parameter $\phi \in [0,1)$, and is defined as follows, 
\begin{align}
P(r) = P(r \mid \sigma, \phi) = \frac{1}{Z} \phi^{d(r,\sigma)}
\end{align}
where $r$ is a ranking, $d$ is a distance function between rankings, and $Z$ is a normalization constant.

For $\phi \rightarrow 1$, the rankings become equiprobable, while $\phi=0$ concentrates all the weight on the central ranking $\sigma$. The most commonly used distance functions are the Kendall Tau and Spearman Footrule distances~\cite{vitelli2018probabilistic}. The Kendall Tau distance counts the number of pairwise disagreements between two rankings, while the Spearman Footrule distance sums the deviations in ranking for each item. We experiment with both in this work and find very similar results, so we report only results with the Kendall Tau distance.

\begin{figure}
    \centering    \includegraphics[width=1\columnwidth]{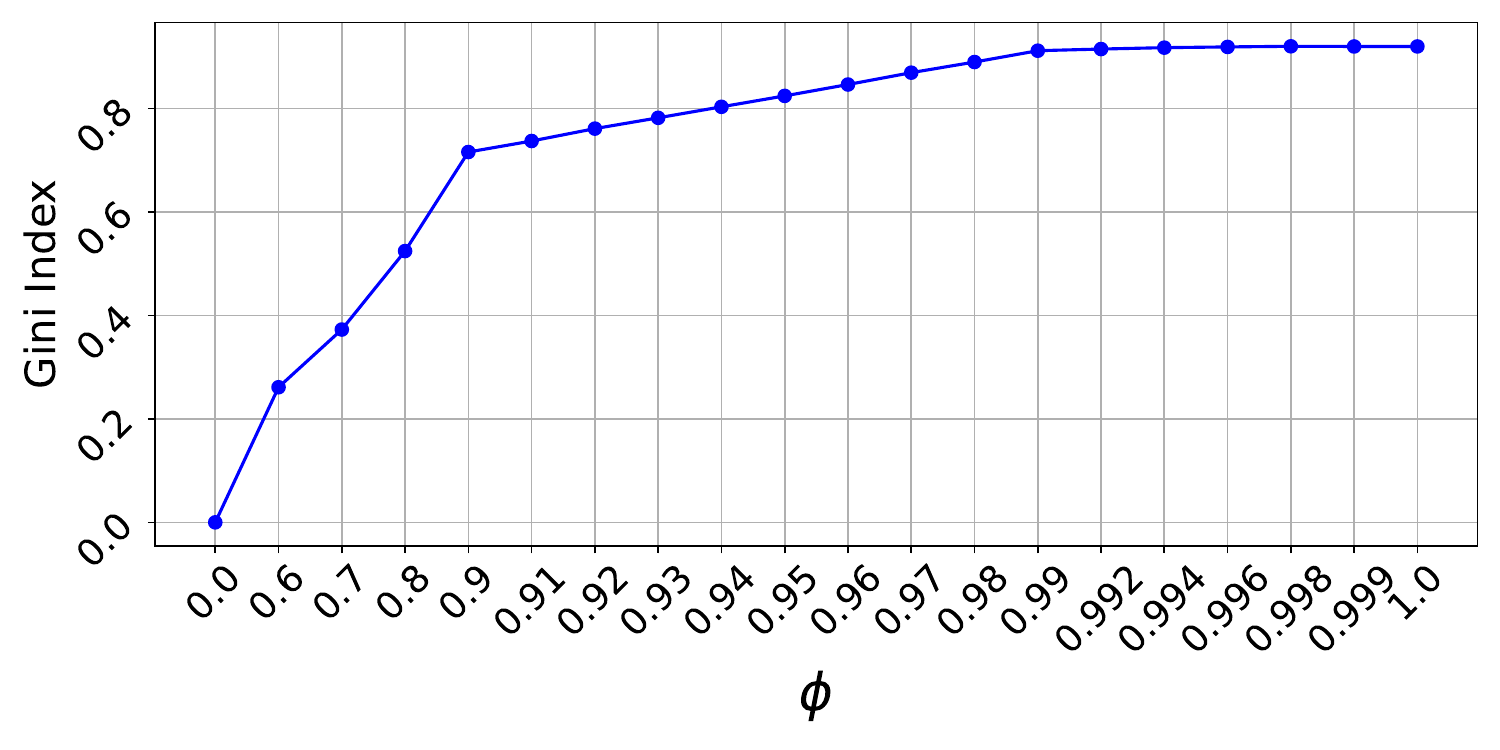}
    \caption{Gini index across neighborhoods increases with the increase in the dispersion parameter$~\phi$}
    \label{fig:gini_vs_phi}
\end{figure}

By using $\sigma = \sigma_0$ and varying the dispersion parameter, we can effectively control how spread out the High-Risk properties are over neighborhoods; we assume that the sampled permutation maintains property identities (and, therefore, risk score/status), while ``moving'' them into different neighborhoods. For a simple example, suppose there are 6 properties in total, spread across 3 neighborhoods, and the ranking $(5, 1, 3, 2, 6, 4)$ is sampled. This means that the properties ranked fifth and first in terms of risk are in Neighborhood 1, the properties ranked third and second are in Neighborhood 2, and the properties ranked sixth and fourth are in Neighborhood 3. In the homogeneous risk ranking, the two riskiest properties are in Neighborhood 1, the next two in Neighborhood 2, and the two least risky in Neighborhood 3.

We note the model above could in theory lead to some permutations that are similar in risk characteristics but with neighborhood ``identities'' changed while appearing to be distant from the baseline rankings. In order to ensure that this does not create an unintended effect, we conducted a simulated experiment with varying values of $\phi$ to ensure that increasing $\phi$ does affect the overall distribution of risk across neighborhoods as intended. The simulation involved $600$ properties divided into $30$ neighborhoods. Keeping the center order ($\sigma_0$) homogeneous, we varied $\phi$ in the range (0,1). We then estimated the Gini index across neighborhoods. Figure~\ref{fig:gini_vs_phi} shows that as the value of the dispersion parameter increases, the Gini index also increases. Thus, increasing $\phi$ does indeed induce a more dispersed distribution of High-Risk properties and reduces homogeneity across neighborhoods.

\subsection{Canvassing Costs and Policies}

The resource allocation task we specifically consider is canvassing high(er)-risk properties to provide information about resources for those who may be facing eviction -- a task that has received prior attention in the literature~\cite{mashiat2024beyond,tabar2022forecasting}. The importance of the neighborhood structure is that the marginal cost of visiting additional properties in a given neighborhood is lower than the cost of going to a different neighborhood. To model this, while keeping the analysis tractable, we fix the cost of traveling between properties within a neighborhood as 1 and allow the cost of traveling between neighborhoods, $\alpha$, to vary.\footnote{Of course, in the real-world context both intra- and inter-neighborhood travel costs will vary, but we find this a useful and reasonable approximation in experiments with actual data.}

\textbf{Non-targeting} policies conduct outreach at the neighborhood level. The structure is simple; policies rank neighborhoods according to the number of High-Risk properties and allocate outreach canvassers to all properties within the neighborhood, continuing down the list of neighborhoods until the budget runs out.

Prediction-based \textbf{targeting} policies use individual risk assessments to decide which properties to canvass. We consider two types of such policies.~\emph{High-Risk Property Targeting (HPT)} is similar to non-targeting policies with the difference that within a neighborhood, the policy only canvasses High-Risk properties. Thus, HPT ranks neighborhoods in terms of the number of High-Risk properties and then sequentially visits all High-Risk properties in each neighborhood, in that order, until exhaustion of the budget.

In some situations, fairness or regulatory constraints may require that higher-risk properties are always canvassed preferentially over lower-risk ones. In these cases, the actual risk score or rank matters, not just whether a property is High- or Low-Risk. We define the \emph{Top-k Property Targeting (TPT)} policy as the one that visits the $k$ highest risk properties. Given a value of $k$, the optimal path through these properties is easy to compute since the properties can be partitioned into their neighborhoods. It is also efficient to perform a binary search to find the highest value of $k$ that can be achieved within a given budget.

\subsection{Evaluation}

In order to evaluate the effectiveness of targeting, we must consider the accuracy of predictions, as well as the budget of the agency implementing the policy. First, we need to consider the relative frequencies of false positives and false negatives in prediction. Let
\begin{align}
    Pr(\text{Eviction}|\text{High-Risk}) = p \nonumber \\
    Pr(\text{Eviction}|\text{Low-Risk}) = q \nonumber
\end{align}

Suppose the agency has a budget of $c$. We define a metric, the \textit{Relative Efficiency of Non-Targeting (RENT)} to evaluate the effectiveness of outreach allocation policies. 
\textit{RENT} at budget $c$ is defined as follows:
\begin{align}
    %RENT_{c} = \frac{|S_{B,c} \cap S_{T,c}|}{|S_{T,c}|}
    \text{RENT}_{c} = \frac{|S_{B,c}|}{|S_{T,c}|}
\end{align}

\begin{figure*}
    \centering
    \captionsetup[subfigure]{justification=centering}
    \subfloat[$N = 50, M = 10$]{\includegraphics[width=0.6\columnwidth]{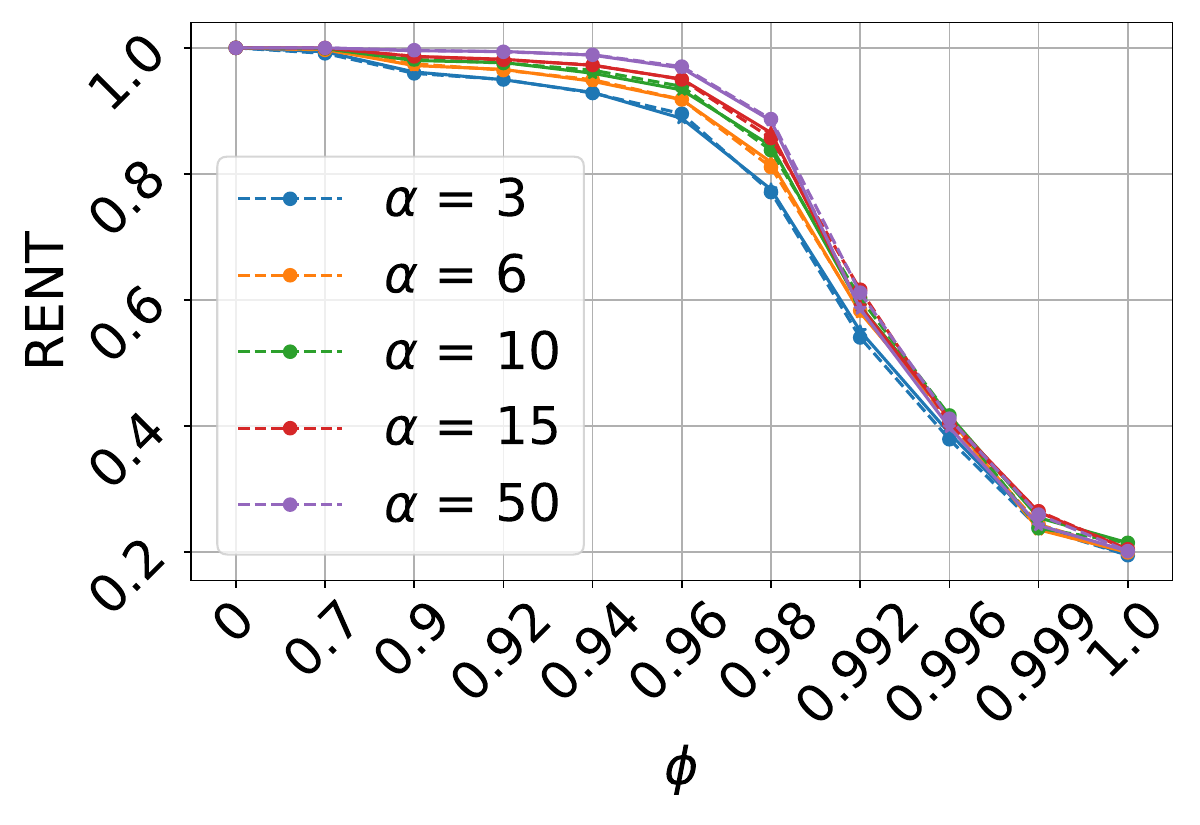}}
    % \hspace{3mm}
    \subfloat[$N = 100, M = 10$]{\includegraphics[width=0.6\columnwidth]{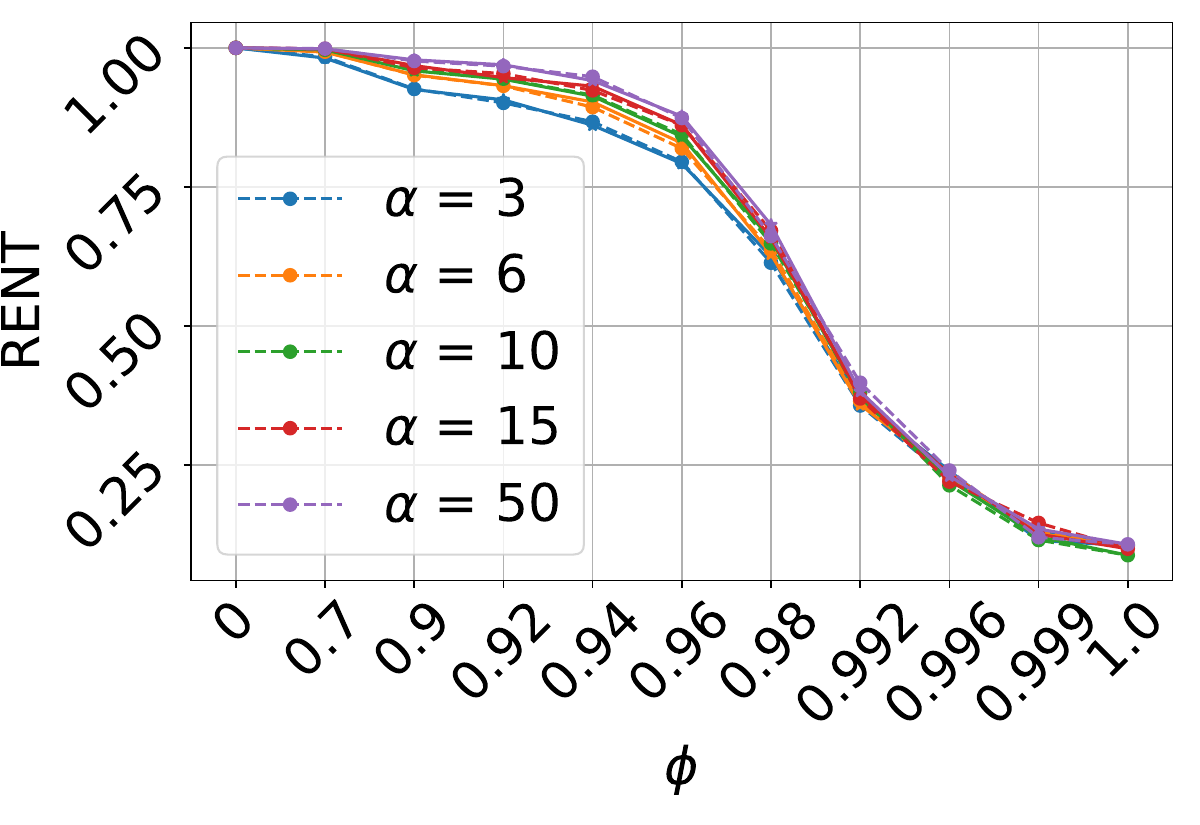}}   
    % \hspace{5mm}
    \subfloat[$N = 150, M = 10$]{\includegraphics[width=0.6\columnwidth]{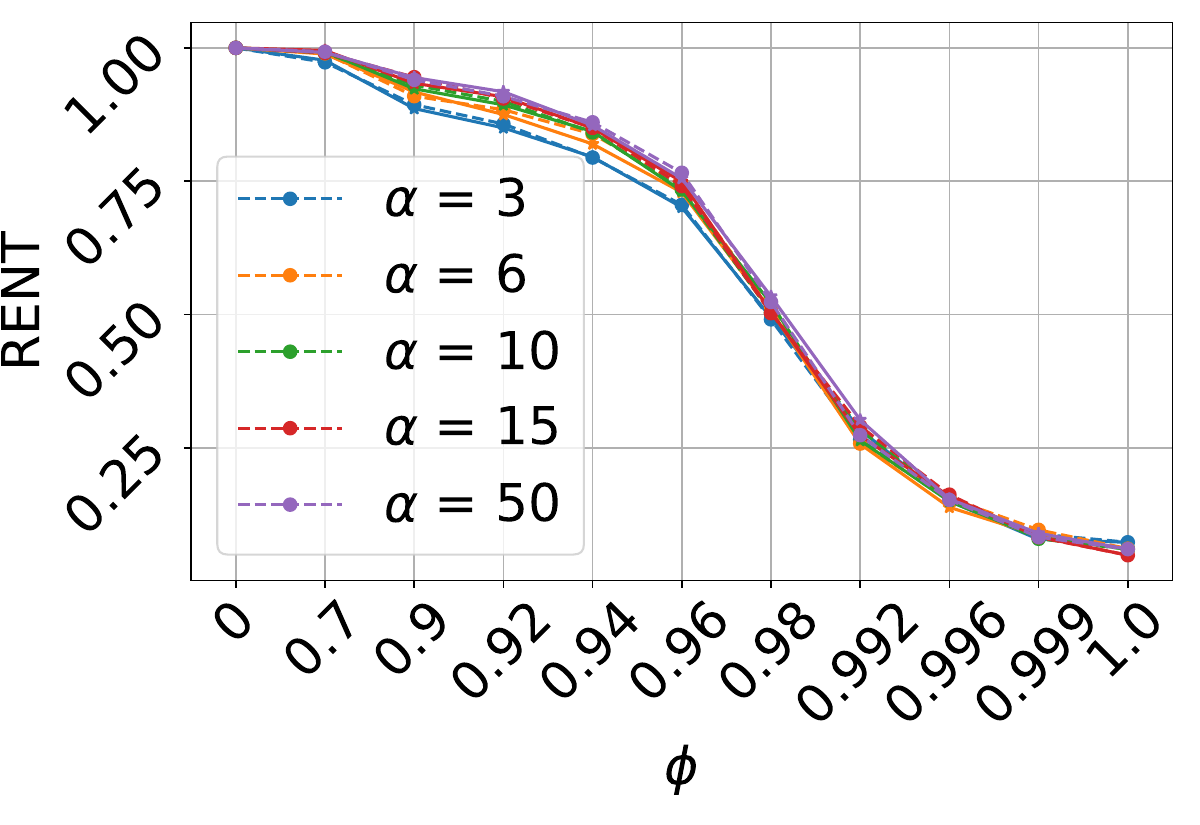}}

    \subfloat[$N = 50, M = 20$]{\includegraphics[width=0.6\columnwidth]{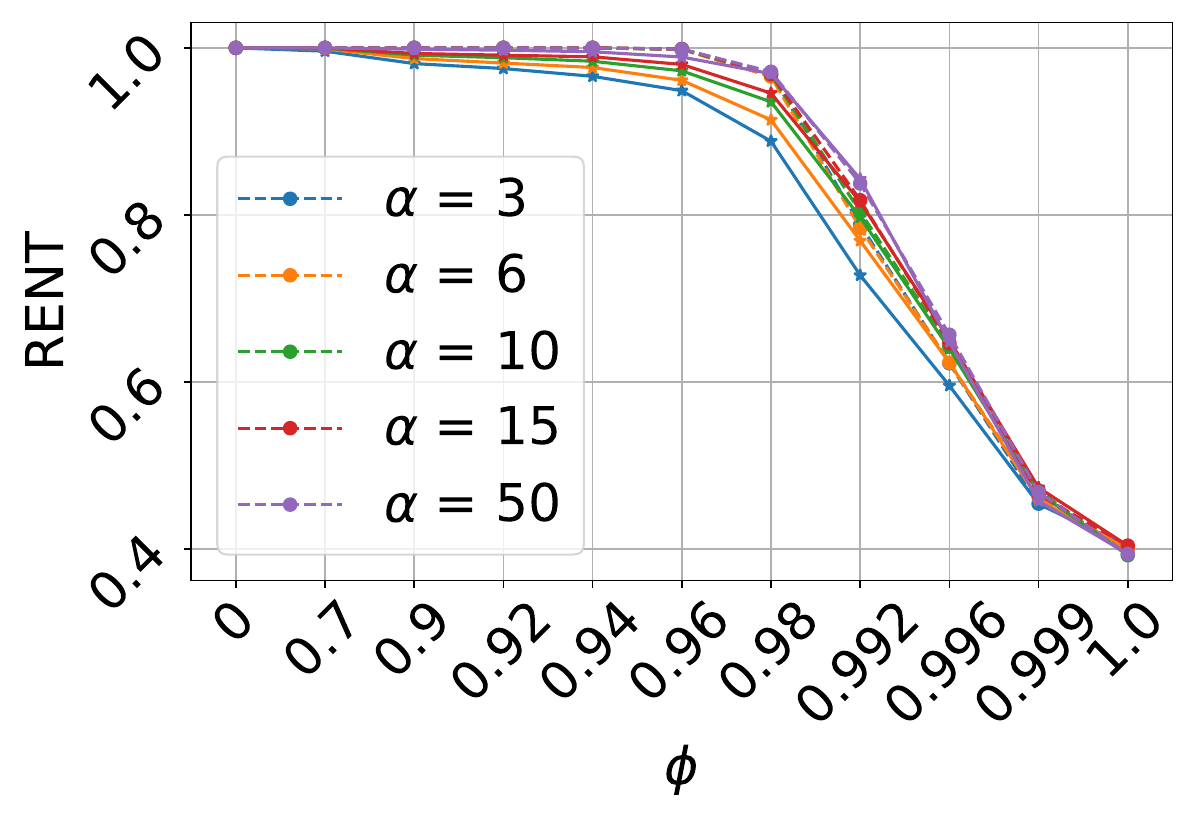}}
    % \hspace{3mm}
    \subfloat[$N = 100, M = 20$]{\includegraphics[width=0.6\columnwidth]{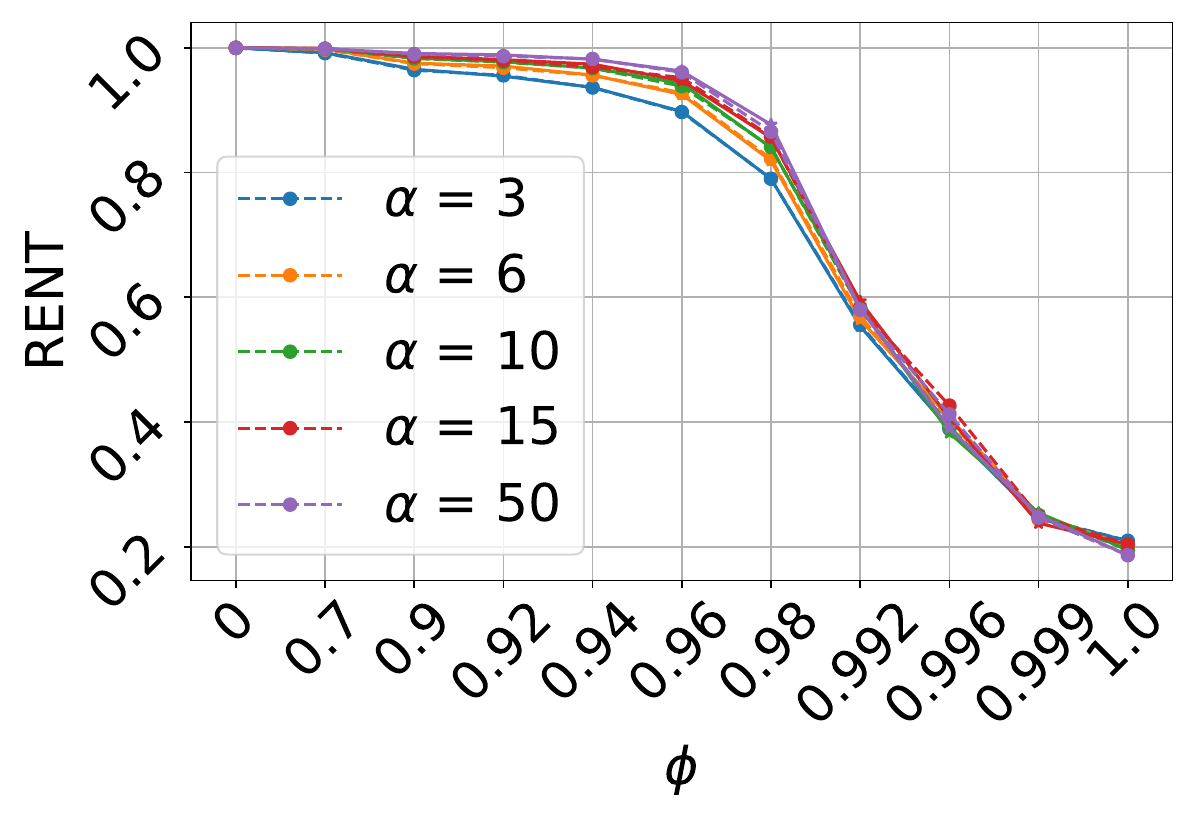}}   
    % \hspace{5mm}
    \subfloat[$N = 150, M = 20$]{\includegraphics[width=0.6\columnwidth]{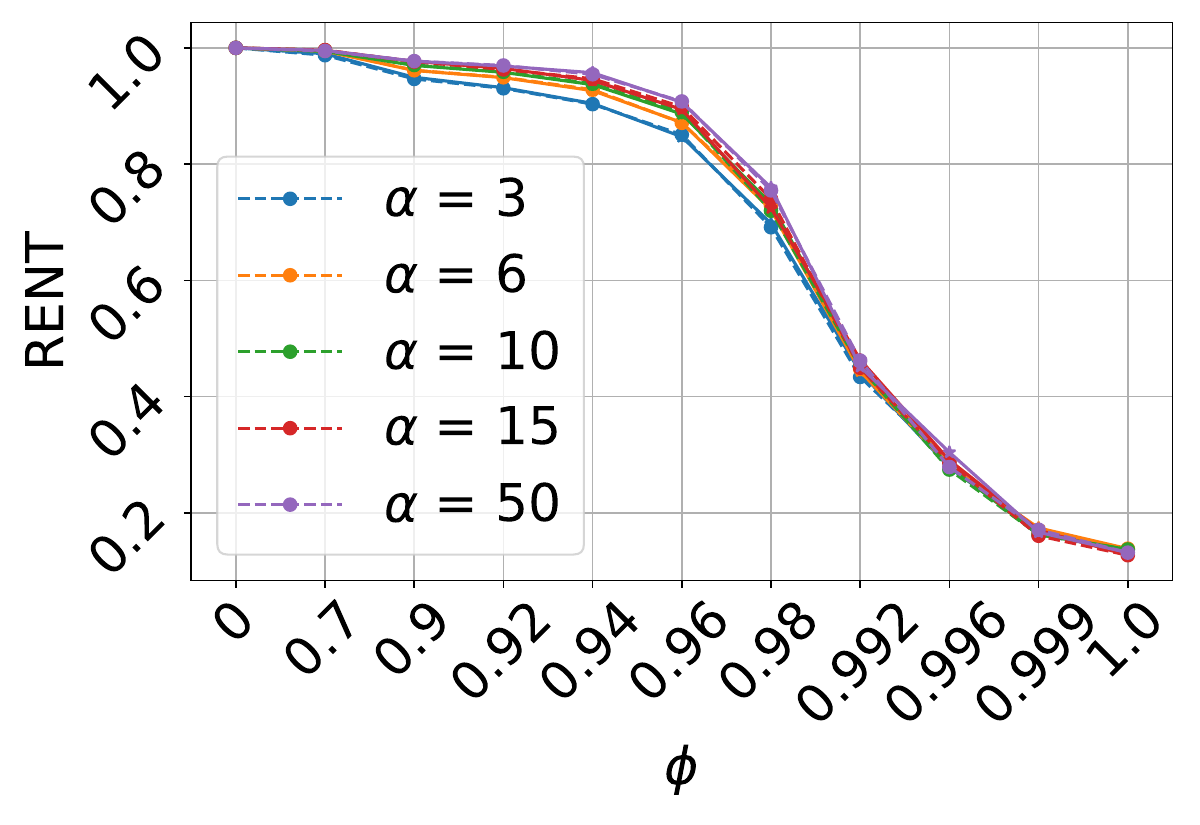}}

    \subfloat[$N = 50, M = 30$]{\includegraphics[width=0.6\columnwidth]{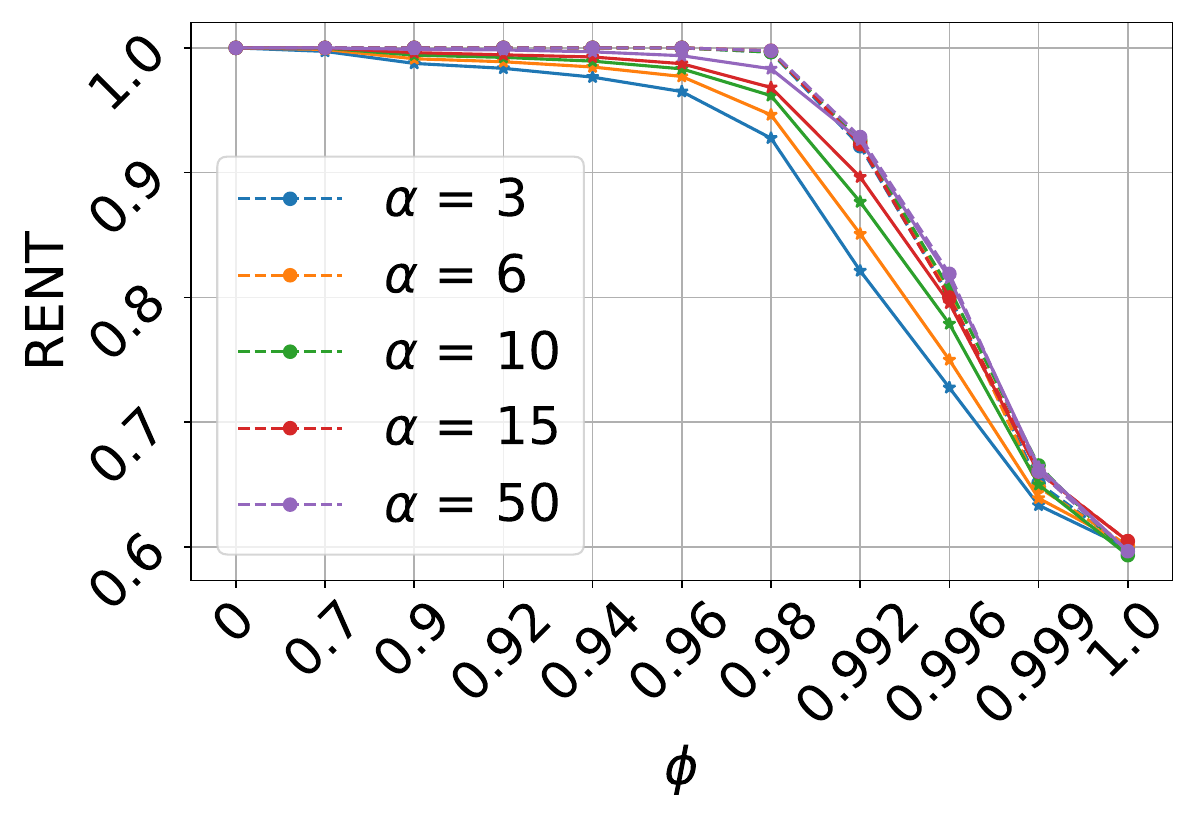}}
    % \hspace{3mm}
    \subfloat[$N = 100, M = 30$]{\includegraphics[width=0.6\columnwidth]{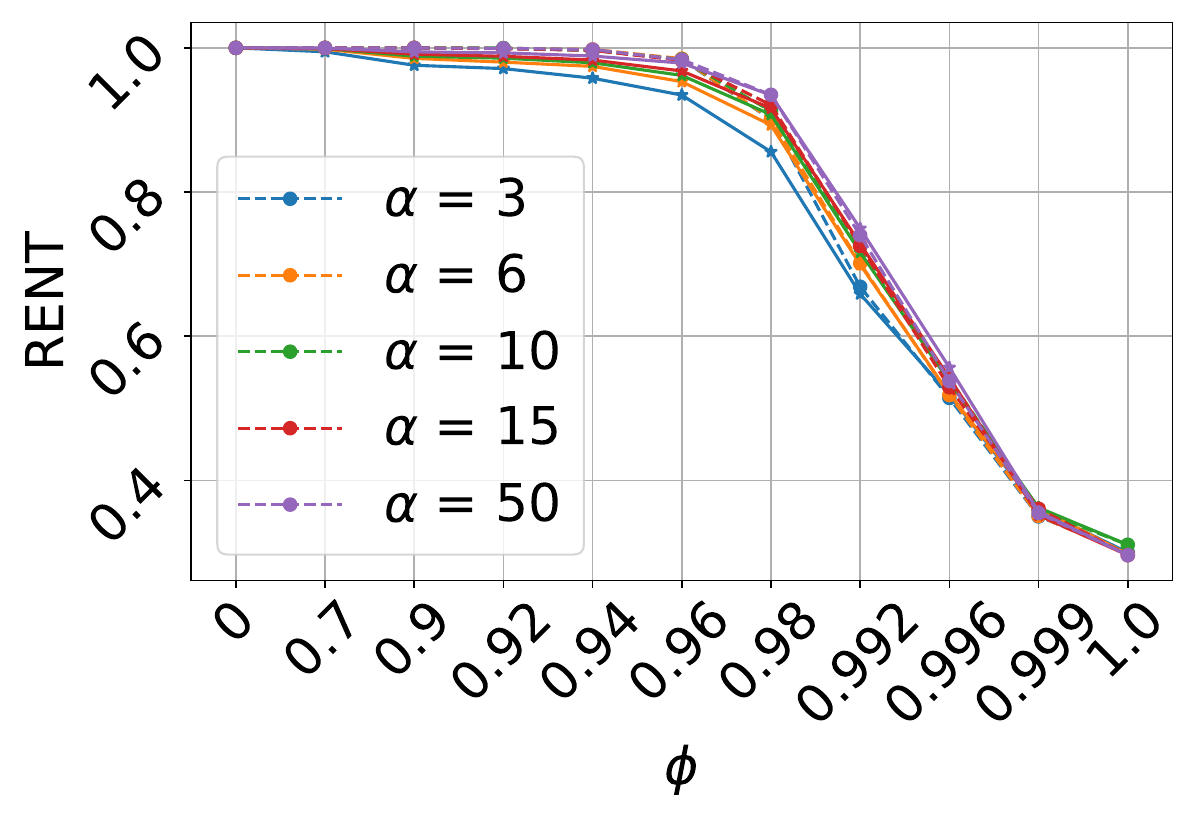}}   
    % \hspace{5mm}
    \subfloat[$N = 150, M = 30$]{\includegraphics[width=0.6\columnwidth]{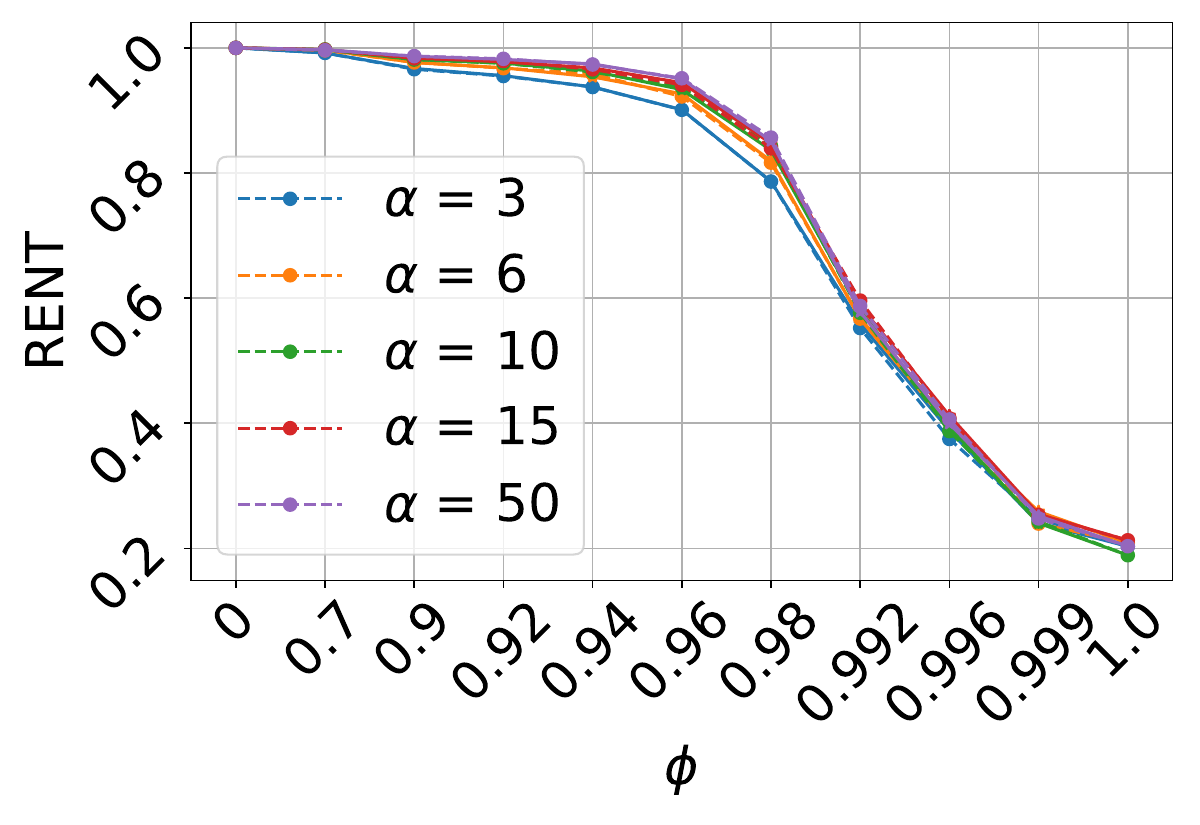}}
    \caption{Results of simulation for $600$ properties located in $N$ neighborhoods. The total cost is estimated for canvassing the top $M$  neighborhoods ordered by the central ranking. $\phi$, the dispersion parameter, is on the X-axis, while the
    Y-axis shows the $RENT$ values. For both \textbf{targeting} policy \textit{HPT} (solid) and \textit{TPT} (dashed), as the distribution of inequality across neighborhoods becomes less homogeneous and divorced from the central order, $\text{RENT}$ 
    decreases. As a result, the effectiveness of \textbf{targeting} policies increases.}
 
    \label{fig:simulation}
    % \Description{}
\end{figure*}
\begin{figure*}
    \centering
    \subfloat[]{\includegraphics[width=0.65\columnwidth]{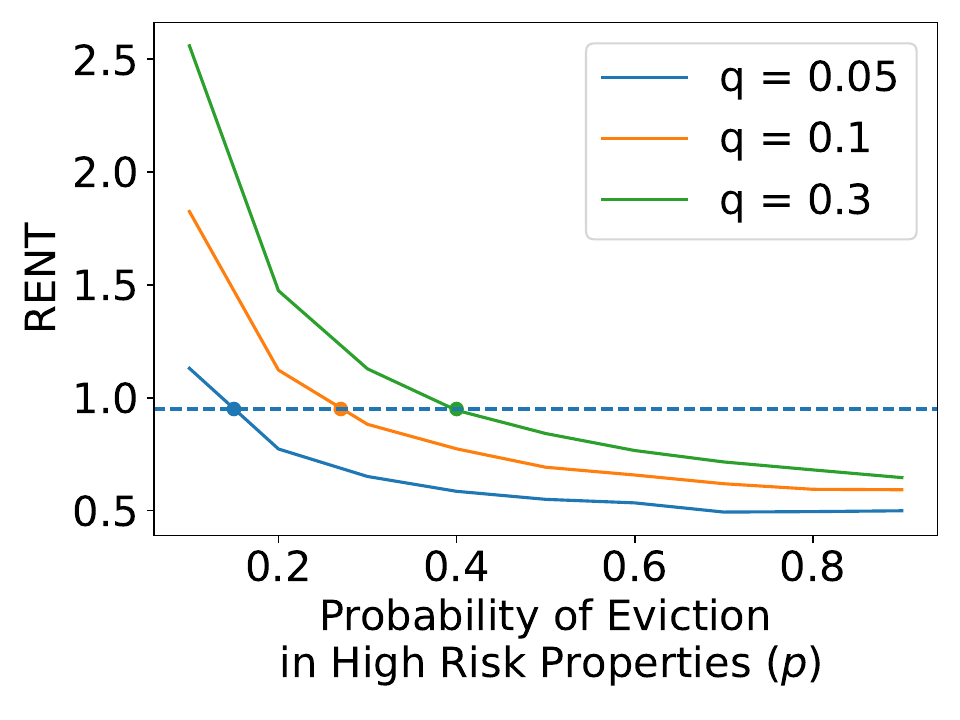}\label{fig:panel_a_q}}  
    \subfloat[]{\includegraphics[width=0.65\columnwidth]{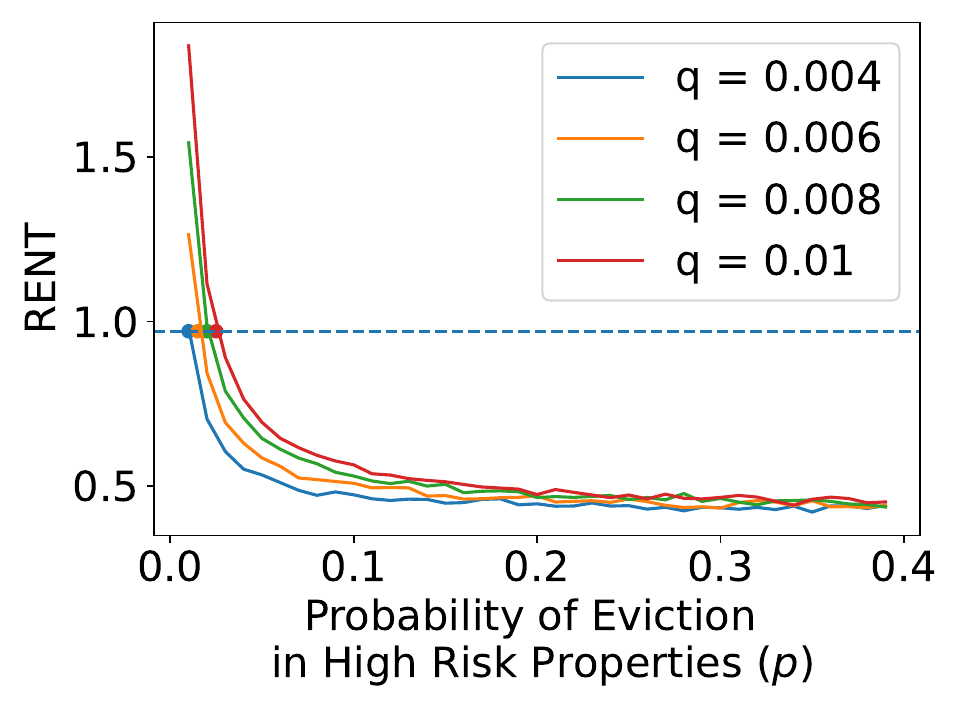}\label{fig:panel_b_q}}

    \caption{RENT as a function of the precision of prediction ($p$) and the false negative rate ($q$). We see that RENT is well below 1, indicating a high relative value of targeting for most reasonable values of precision and false negative rate. Figure (a) shows a broader set of possible values, while Figure (b) focuses on values of $p$ and $q$ that are more plausible for eviction prediction, where the base rate of actual evictions is quite low, and therefore precision is unlikely to exceed $0.4$, while the false negative rate will usually be below $0.01$.}

    \label{fig:simulation_likely_hood}
    % \Description{}
\end{figure*}

Where, $S_{B,c}$ denotes the set of properties that experienced eviction canvassed by the \textbf{non-targeting} policy and $S_{T,c}$ denotes the set of properties that experienced eviction canvassed by a \textbf{targeting} policy for budget $c$. The lower the RENT, the more effective targeting is.

\section{Simulation Results}

We now turn to understanding how the dispersion parameter $\phi$ affects $\text{RENT}$. We do so by simulating a city with different numbers of neighborhoods. Suppose that there are $600$ properties in a city. As case studies, we divide these properties into $N = \{50,100,150\}$ neighborhoods. 

We consider values of $\alpha$ (the travel cost between neighborhoods) of $\{3,6,10,15,50\}$. For each value of the dispersion parameter $\phi$ examined, we sample rankings from a Mallows model with parameters $\sigma_{0}$ and $\phi$.

We determine a budget $c$ by calculating the cost to canvass all properties within $M$ neighborhoods using a non-targeting policy. Subsequently, we identify the properties canvassed by HPT and TPT at budget $c$. We assume that in HPT, the top $20\%$ of all properties are categorized as High-Risk, and in TPT the value of $k$ is set to the maximum number of highest ranked properties that can be visited within budget $c$.
For our initial experiments, we assume that $p=1$ and $q=0$, so that the risk predictions are essentially perfect. Figure~\ref{fig:simulation} shows how $\text{RENT}$ varies with $\phi$ for different settings of $N, M,$ and $\alpha$.

There are two main observations from these results. First, the RENT metric is close to 1 for a wide range of $\phi$ values and then drops rapidly as $\phi$ approaches 1. This finding implies that targeting is much more valuable when High-Risk properties are widely dispersed between neighborhoods, as expected. It raises the question of how the $\phi$ parameter corresponds to what we might expect to see in the real world, which we address in the next section with data from St. Louis. The second observation is that the relative value of targeting is higher when there are more neighborhoods and lower travel costs between neighborhoods. Although intuitive, our framework provides a way of capturing, formalizing, and qualifying the results.

The next question concerns how the accuracy of predictions affects RENT. As $p$ declines and $q$ increases, the relative value of targeting must decrease. It is easy to compute values of $S_T$ and $S_B$ because $S_T$ in expectation is just $p$ times the number of properties that the targeting method visits, while $S_B$ is $p$ times the number of high-risk properties visited by the non-targeting method plus $q$ times the number of low-risk properties visited.

Figure~\ref{fig:simulation_likely_hood}(a) shows how RENT varies for $p = [0.1 - 0.9]$, $q=[0.05,0.1,0.3]$, and $\phi = 0.99$ (close to the $\phi$ values for real eviction data, later discussed). Targeting performs better as long as the precision $p$ is reasonably higher than the false negative rate $q$ (for example, once $p = 0.3$ for $q = 0.1$ and about $p = 0.15$ for $q = 0.05$). Since evictions are relatively low baseline probability events, we actually typically see very low values of both $p$ (less than $0.4$) and $q$ (less than $0.01$) in eviction prediction, while still being able to maintain a predictive edge. Figure~\ref{fig:simulation_likely_hood}(b) shows the RENT metric in this regime, and it is interesting to see that RENT is quite low, implying significant benefits of targeting when $p$ is $0.05$ or above.

\begin{figure}
    \centering
    \subfloat[Residential properties (properties with higher eviction risk are marked red and low risk are marked blue).]{
    \includegraphics[width=1\columnwidth]{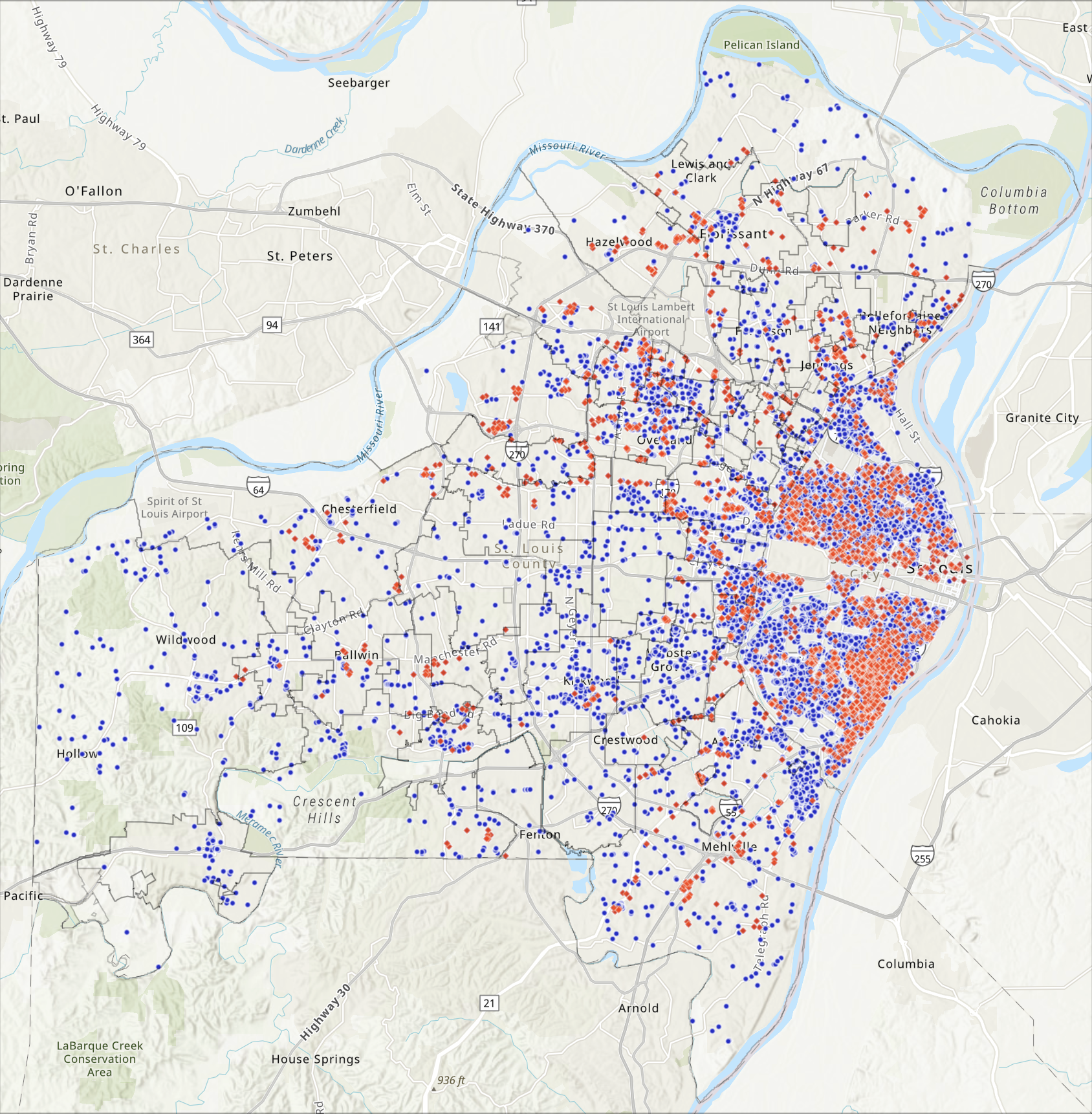}\label{fig:neigh_dist}}
    
    \subfloat[ Distribution of High-Risk Rate]{
    \includegraphics[width=1\columnwidth]{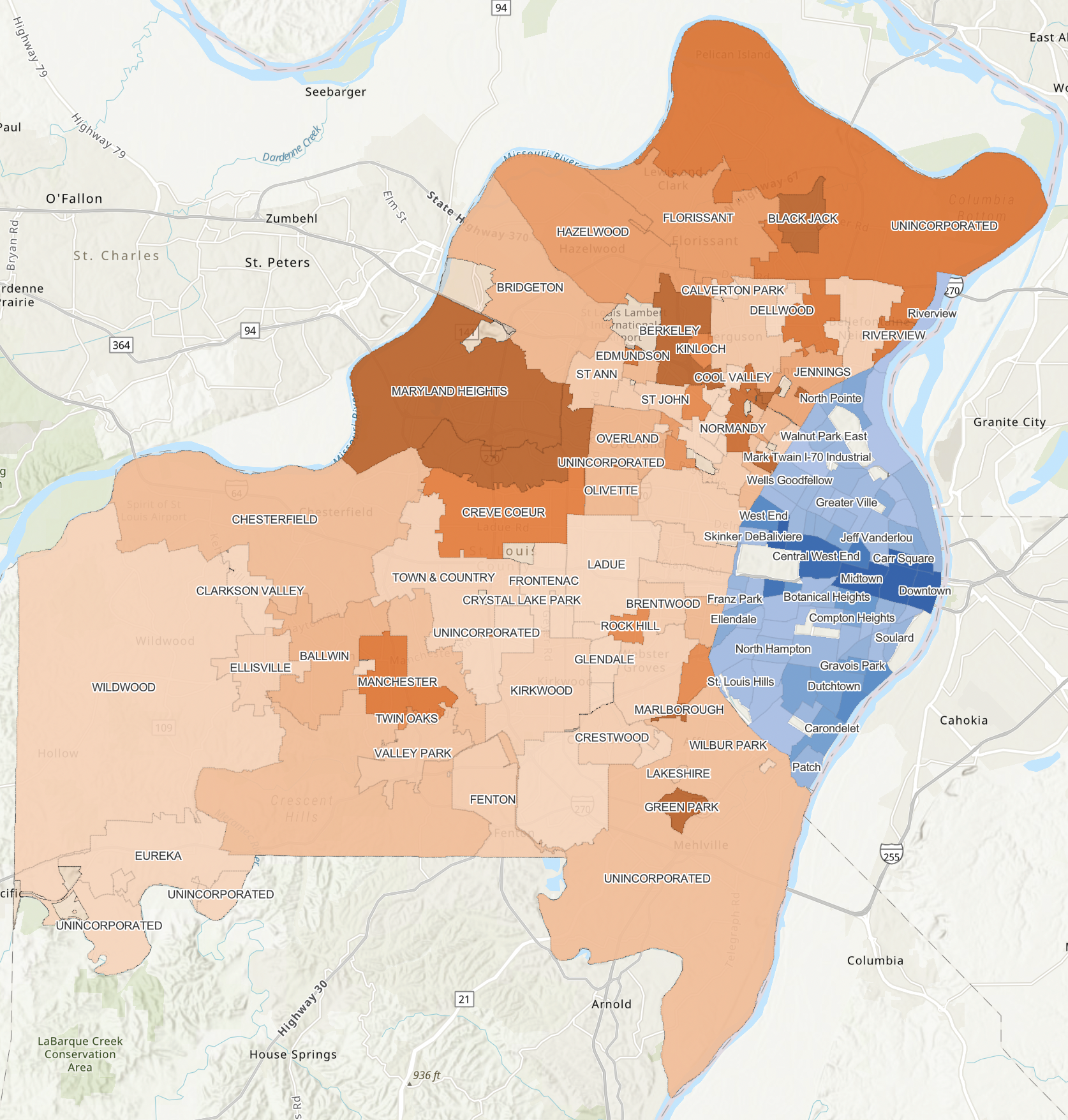}\label{fig:mun_dist}}
    \caption{(a) Residential properties in St. Louis city and county; the red dots represent the High-Risk properties, and the blue dots represent Low-Risk properties (b) Distribution of High-Risk Rate in St. Louis City (blue) and County (orange) neighborhoods; the color scale represents the value of the rate, and darker shades indicate higher values.}
    \label{fig:st_louis_count}
\end{figure}

\section{Application to Real-World Data}

In order to understand the applicability of our framework, we turn to a detailed analysis of data on eviction risk from St. Louis, MO. We use geospatial data on properties, predictions of eviction risk, and actual court eviction filings from recent work of ~\citet{mashiat2024beyond}. St. Louis is a useful example because it is generally considered highly segregated. St. Louis City is ranked 10th in segregation among US cities according to 2020 Census data by the Berkeley Othering and Belonging Institute, and the region is infamous for the so-called ``Delmar Divide'' where areas north and south of Delmar Boulevard exhibit significantly different racial and socioeconomic makeups and corresponding disparities in life expectancy. We consider census-defined neighborhoods of St. Louis City and municipalities of St. Louis County but refer to these units simply as neighborhoods in what follows for the sake of simplicity. 

\begin{figure}
    \centering
    \includegraphics[width=0.9\columnwidth]{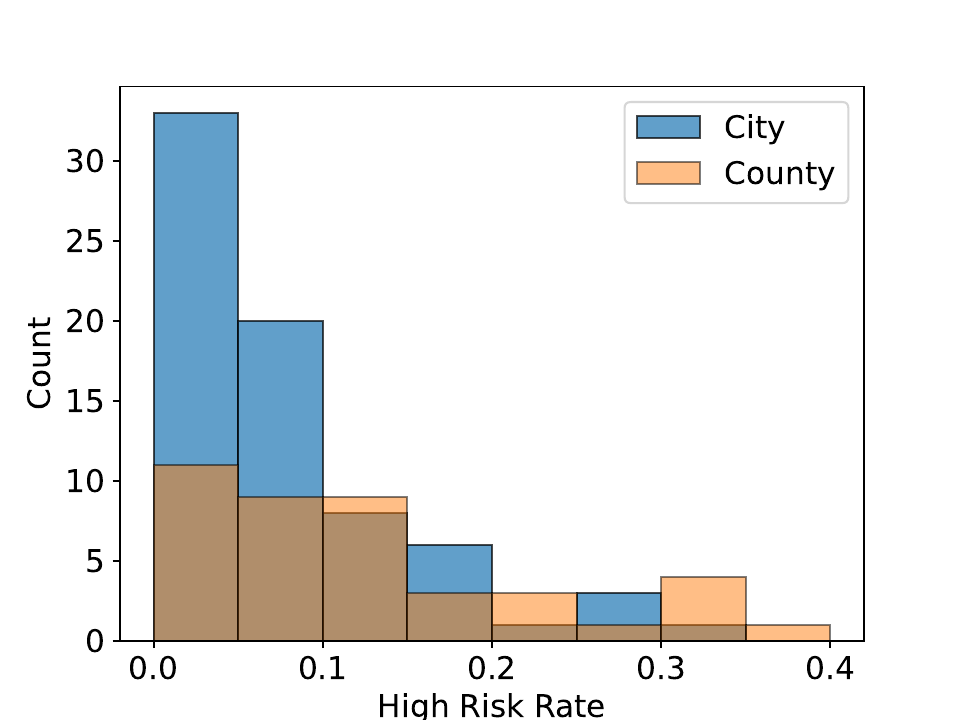}
    \caption{Distribution of High-Risk Rate in St. Louis City (blue) and County (orange) neighborhoods.}
    \label{fig:risk_score_distribution}
\end{figure}

\subsection{Dataset Description and Background}

The eviction dataset provided by Mashiat et al. has eviction records and risk scores for $26770$ residential properties. Among these $19232$ are located in St. Louis City and $7543$ in St. Louis County~\cite{mashiat2024beyond}.

St. Louis City has $79$ neighborhoods, while the County has $98$ neighborhoods. To remove outliers and capture distributional trends, we only consider neighborhoods with more than $30$ residential properties with two or more units.

Mashiat et. al. categorized $8.5\%$ of the properties as High-Risk in their best-performing eviction prediction model, where $1434$ and $864$ properties are located in the City and the County, respectively. We compute the ratio of High-Risk properties to the total number of properties within each neighborhood and call this the \textit{High-Risk Rate}. Figure~\ref{fig:mun_dist} and~\ref{fig:risk_score_distribution} show the distribution of the High-Risk Rate across neighborhoods. The High-Risk properties are also marked as red in~Figure~\ref{fig:neigh_dist} and are stratified across the City and the County. We can see that the distribution of the High-Risk Rate across neighborhoods is quite different between the City and the County, with the County displaying much more uniformity in the rate of High-Risk properties across neighborhoods~(Figure~\ref{fig:risk_score_distribution}).

We now turn to calibrating the model developed above to the data from St. Louis City and County in order to find values of $\phi$ consistent with the observed distribution of High-Risk properties.

\subsection{Calibrating the Mallows Model}

Since our interest is in modeling the heterogeneity of the rates of High-Risk properties across neighborhoods, we conduct two types of calibration processes. For both of them, we simplify the process by assuming that all neighborhoods have $n = 100$ properties each.
After removing outliers and neighborhoods with fewer than $30$ properties, the remaining number of neighborhoods ($N$) is $62$ for the City and $43$ for the County.
%For the city, $N = 62$, for the county, $N = 43$. 

In our first process, we set the High-Risk rates for the City to $0.07$ and the County to $0.11$, based on real-world data. We initialize our homogeneous central ranking $\sigma_{0}$ such that the top $7\%$ and $11\%$ of properties are categorized as High-Risk for the City and the County, respectively. We perform $1000$ simulations for a range of dispersion parameters $\phi$. In each simulation, we rank the neighborhoods by their risk rates. We record the minimum and maximum High-Risk rates for each neighborhood by rank. This gives a plausible distribution for what the High-Risk Rate could be in the neighborhood ranked $r$ for each value of $r$. We then compare the observed distributions of the High-Risk rates by rank with the ranges generated in simulation for different values of $\phi$. We find that the risk rate of each neighborhood falls within the risk ranges of the generated rankings produced by the framework when $\phi = \{0.999 \text { (City)}, 0.9867 \text{ (County)}\}$. 

\begin{figure}
    \centering
    \includegraphics[width= 1\columnwidth]{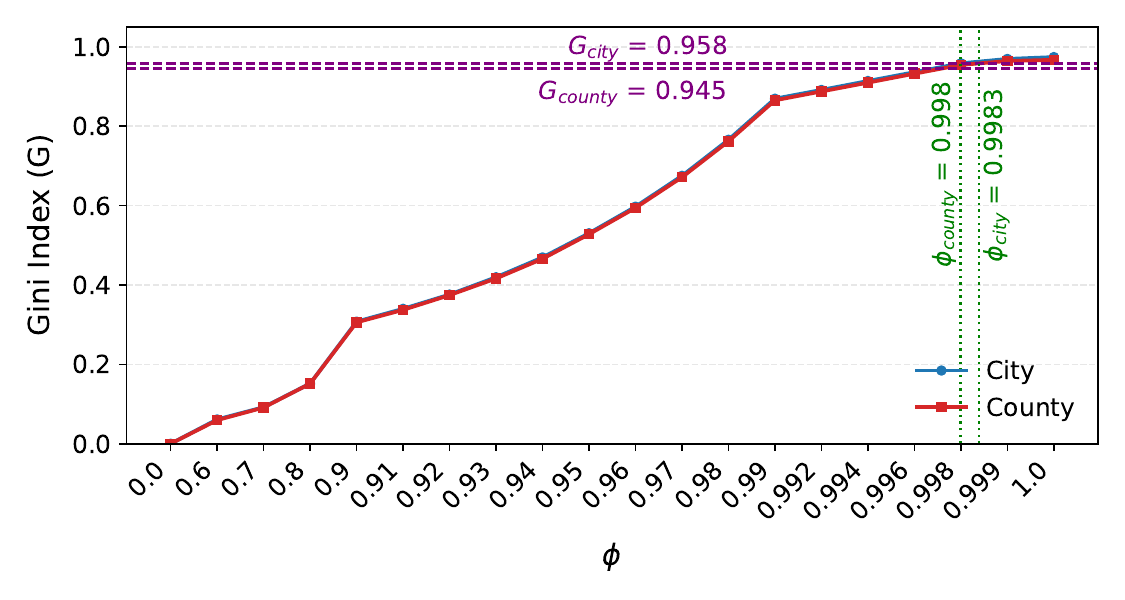}
    \caption{The computed Gini index (G) for neighborhoods in St. Louis city and county aligns with the Gini index corresponding to the dispersion parameter ($\phi$), when $\phi_{city} = 0.9983$ and $\phi_{county}=0.998$. } 
    \label{fig:gini_phi_city_county}
\end{figure}

\begin{figure*}
    \centering
    \subfloat[St. Louis City]{    \includegraphics[width=.85\columnwidth]{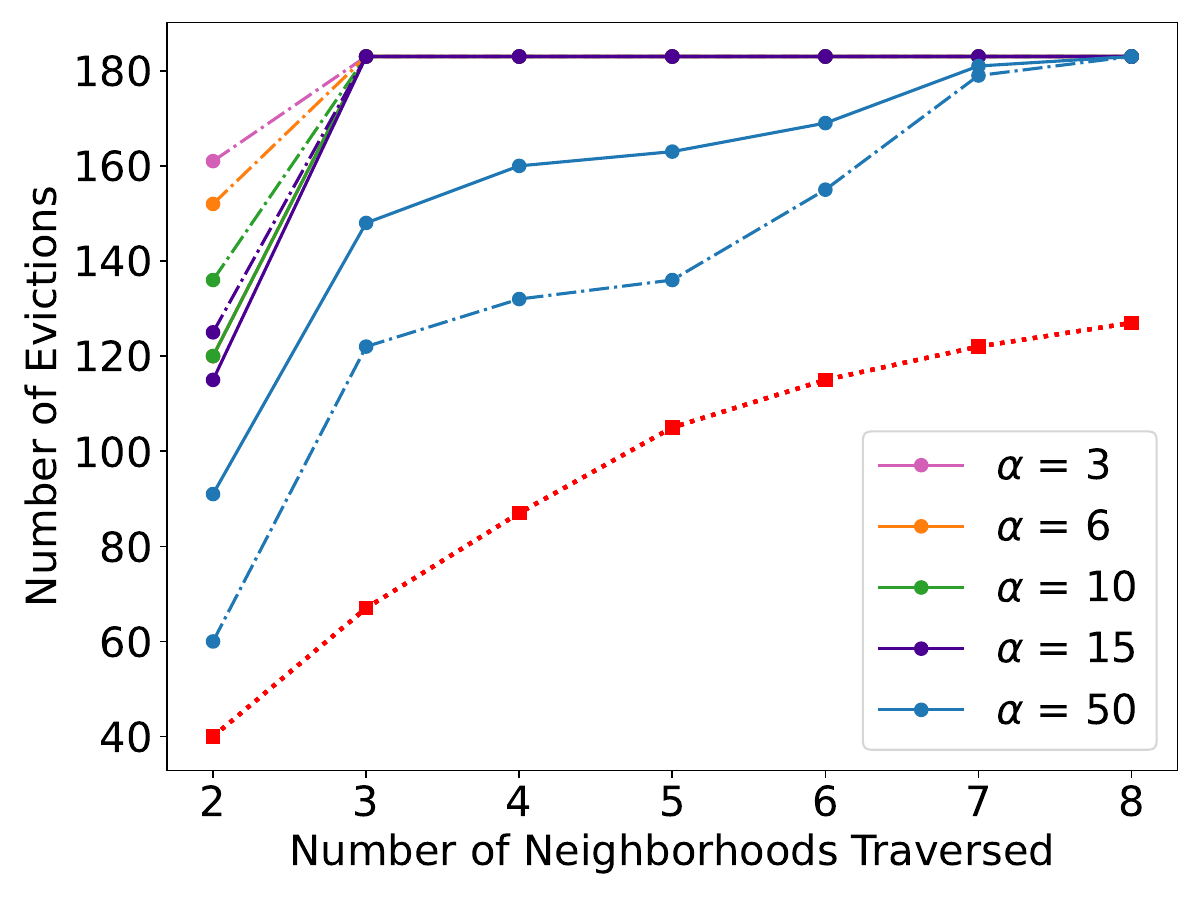}} \hspace{2mm}   
    \subfloat[St. Louis County]{\includegraphics[width=.85\columnwidth]{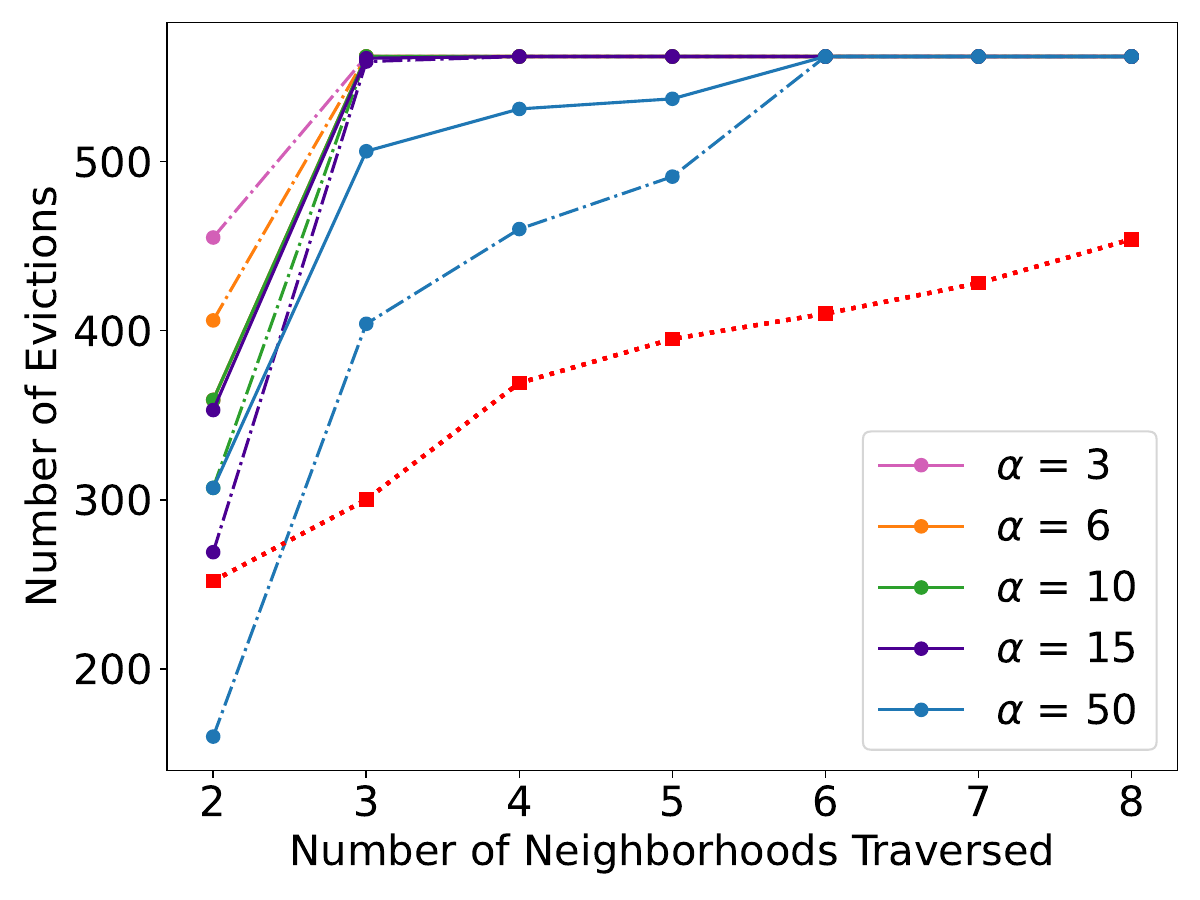}\label{fig:evic_count_county}}

    \caption{ Comparative analysis of HPT (solid) and TPT (dashed) and non-targeting (red line) policies based on the number of discovered evictions in November, December 2022 - January 2023 for St. Louis City and County. The X-axis represents the number of neighborhoods traversed by the non-targeting policy, and the Y-axis represents the number of discovered evictions.}
    \label{fig:num_of_eviction}
\end{figure*}
\begin{figure}
    \centering
    \subfloat[St. Louis City]{   \includegraphics[width=0.85\columnwidth]{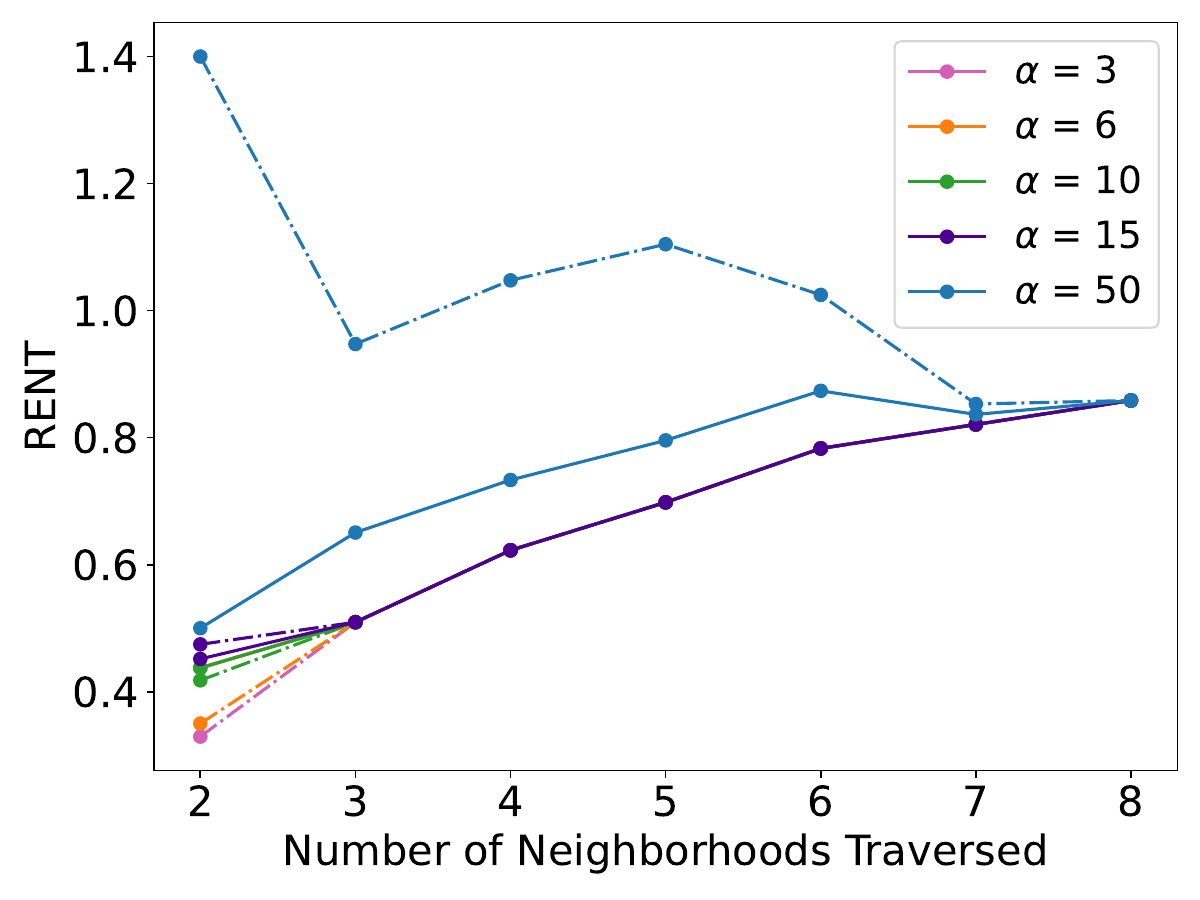}} 
    
    \subfloat[St. Louis County]{\includegraphics[width=0.85\columnwidth]{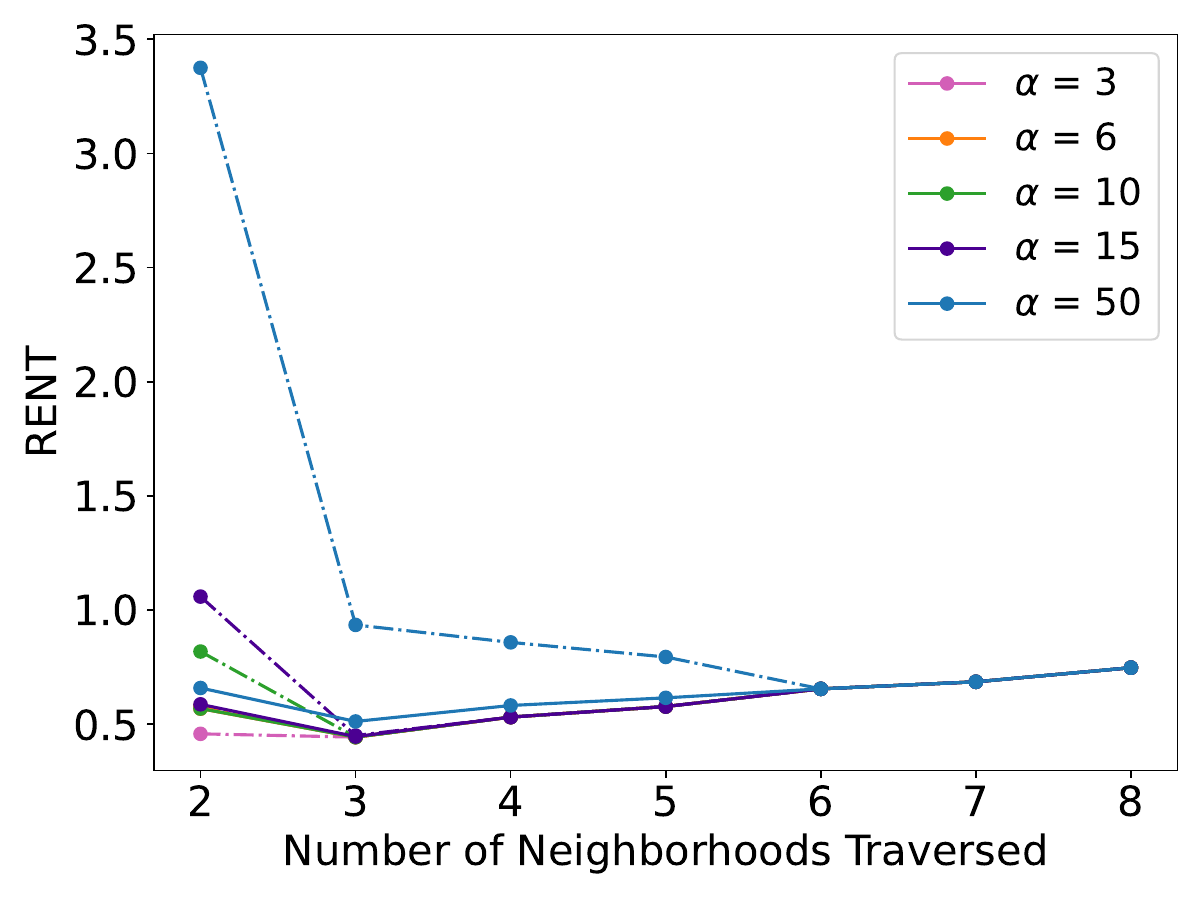}}

    \caption{Comparative analysis of RENT values for HPT (solid) and TPT (dashed) relative to the non-targeting policy for St. Louis City and St. Louis County. The X-axis represents the number of neighborhoods traversed by the non-targeting policy, and the Y-axis represents the RENT metric. }
    \label{fig:comparison_rent}
\end{figure}

In our second process, we estimate the Gini index ($G$) across the same range of dispersion parameters ($\phi$). Figure~\ref{fig:gini_phi_city_county} shows $\phi$ and the corresponding $G$ values. We calculate $G$ based on the distribution of High-Risk properties across neighborhoods in the city ($G_{city} = 0.9583$) and the county ($G_{county} = 0.9453$). We observe from Figure~\ref{fig:gini_phi_city_county}, that the corresponding $\phi$ values for these Gini indices, are $\phi = \{0.9983 ~\text{(City)}, 0.998 \text{ (County)}\}$

The calculated $\phi$ values for both city and county are close to $0.99$ in both analyses. This shows that even for an area considered emblematic of high levels of segregation and inequality, the dispersion parameters in the Mallows model need to be very high to match the characteristics of the real data. These parameter values correspond to regimes where the RENT metric is very low (Figure~\ref{fig:simulation}), indicating that targeting can potentially have significant value.

\subsection{The Value of Targeting}

Next, we turn to comparing the proposed targeting policies (HPT and TPT) relative to the non-targeting policy. As the City and the County are separate jurisdictions, we assume that the budget allocated for outreach in these two areas is determined independently. Suppose that the budgets for canvassing City and County properties are set to $c_{1}$ and $c_{2}$, respectively.

We remove the constraint that mandates an equal number of properties in all neighborhoods and consider the actual number of properties for each neighborhood. We use the risk scores computed for predicting evictions in November, December 2022 - January 2023. We note that the computed values of $p$ and $q$ for the City are $0.078$ and $0.008$, respectively, while those for the County are $0.247$ and $0.005$.

Using prior eviction data, we rank the neighborhoods based on the number of observed evictions in the previous three months (August - October 2022). The cost of traveling between properties is set to 1, while the cost of traveling between neighborhoods $\alpha$ varies, similar to the simulations above.

Figure~\ref{fig:num_of_eviction} shows the comparative analysis between \textbf{targeting}  and \textbf{non-targeting} allocation policies based on the number of discovered evictions. The only situation where the targeting policies do not substantially dominate the non-targeting policy occurs in St. Louis County when the overall budget is low and the travel cost between neighborhoods is extremely high (Figure~\ref{fig:evic_count_county}, $\alpha = 50$) in comparison to the travel cost within a neighborhood, and the targeting policy is constrained to have to visit higher-risk properties first (TPT, rather than HPT). Figure~\ref{fig:comparison_rent} shows the same results in terms of the RENT metric.

\begin{figure}
    \centering
    \includegraphics[width=0.9\columnwidth]{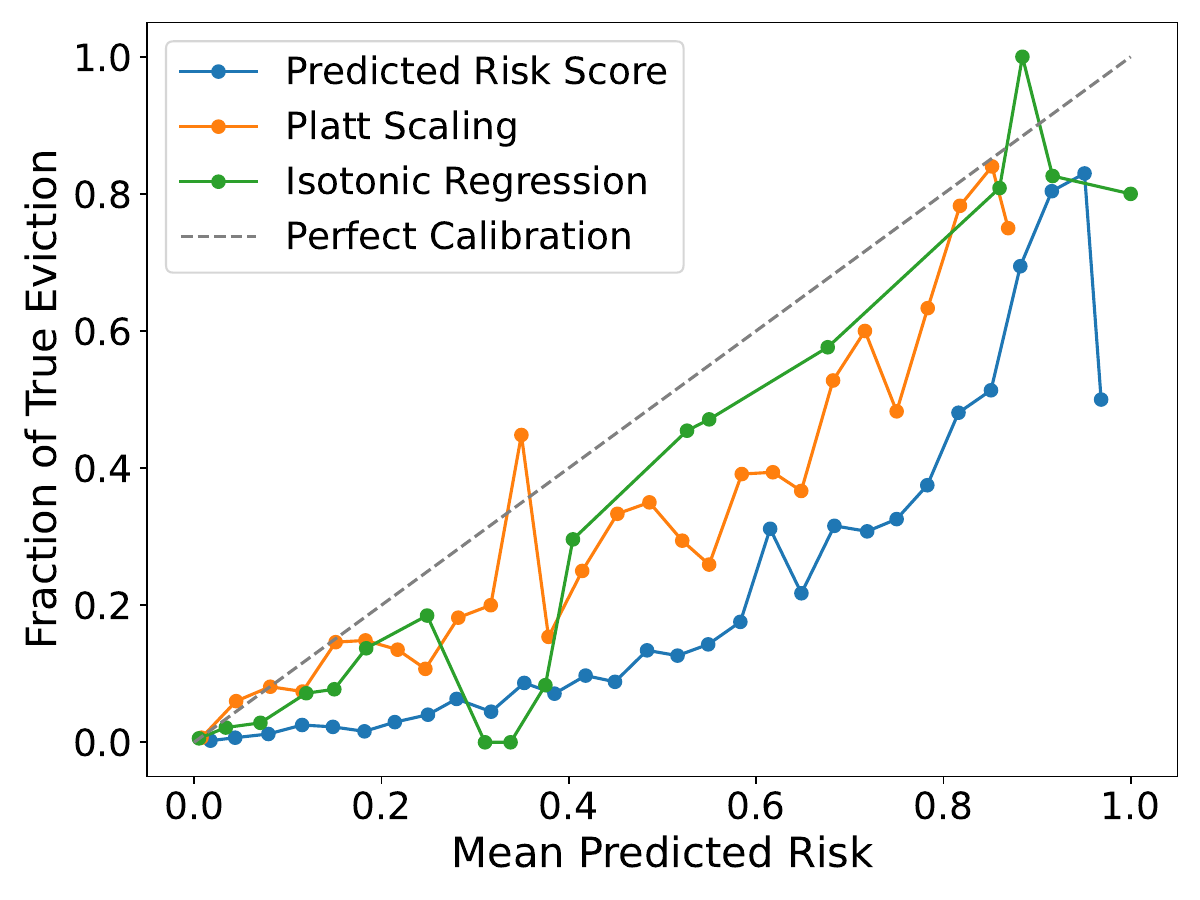}
    \caption{Calibration Curve for eviction risk scores where bin size is 30.} 
    \label{fig:calibration_curve}
\end{figure}

\subsection{The Utility of Targeting}
We now turn to the problem of understanding the value of targeting in terms of eviction reduction under different assumptions about the effectiveness of the intervention that canvassers offer. In order to do this, we first turn to addressing the question of risk score reliability.

\paragraph{Calibrating Risk Scores}To understand the reliability of generated risk scores, we calibrate the risk scores for all properties using both Platt Scaling and Isotonic Regression. Figure~\ref{fig:calibration_curve} shows the calibration curves for the uncalibrated risk scores and the calibrated risk scores. The calibrated risk scores from Isotonic Regression are closer to the line representing perfect calibration. 
We use the eviction probabilities obtained by calibrating risk scores using Isotonic Regression to evaluate and compare the effectiveness of targeting versus non-targeting canvassing policies.

We assume that the intervention reduces the probability of an eviction at a canvassed property by some percentage.
We examine four scenarios: no intervention (where risk scores remain unchanged and no reduction in evictions is assumed), and reduction in eviction probability by 30\%, 50\%, and 70\%.
Our utility then becomes the expected decrease in the number of evictions through the canvassing intervention.

Similar to our earlier experiment, we consider the High-Risk and Low-Risk properties of neighborhoods from both the city and the county. We set the cost based on the number of neighborhoods canvassed. For simplicity, we set the cost of traveling between properties to 1, within a neighborhood, while the cost of traveling between neighborhoods, $\alpha = 3$.

\begin{figure}[ht]
    \centering

    \subfloat[St. Louis City]{   \includegraphics[width=0.825\columnwidth]{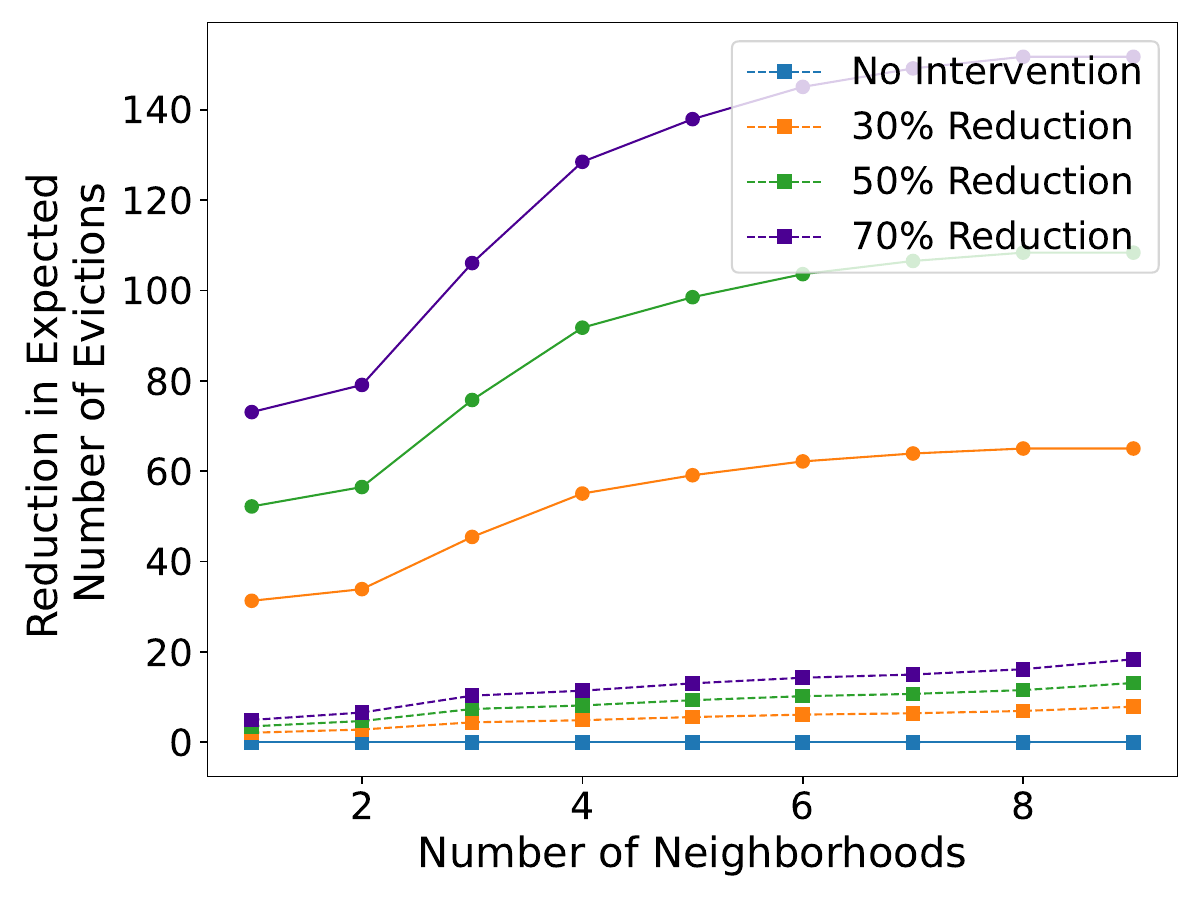}\label{fig:exp_evic_city}} 
    
    \subfloat[St. Louis County]{\includegraphics[width=0.825\columnwidth]{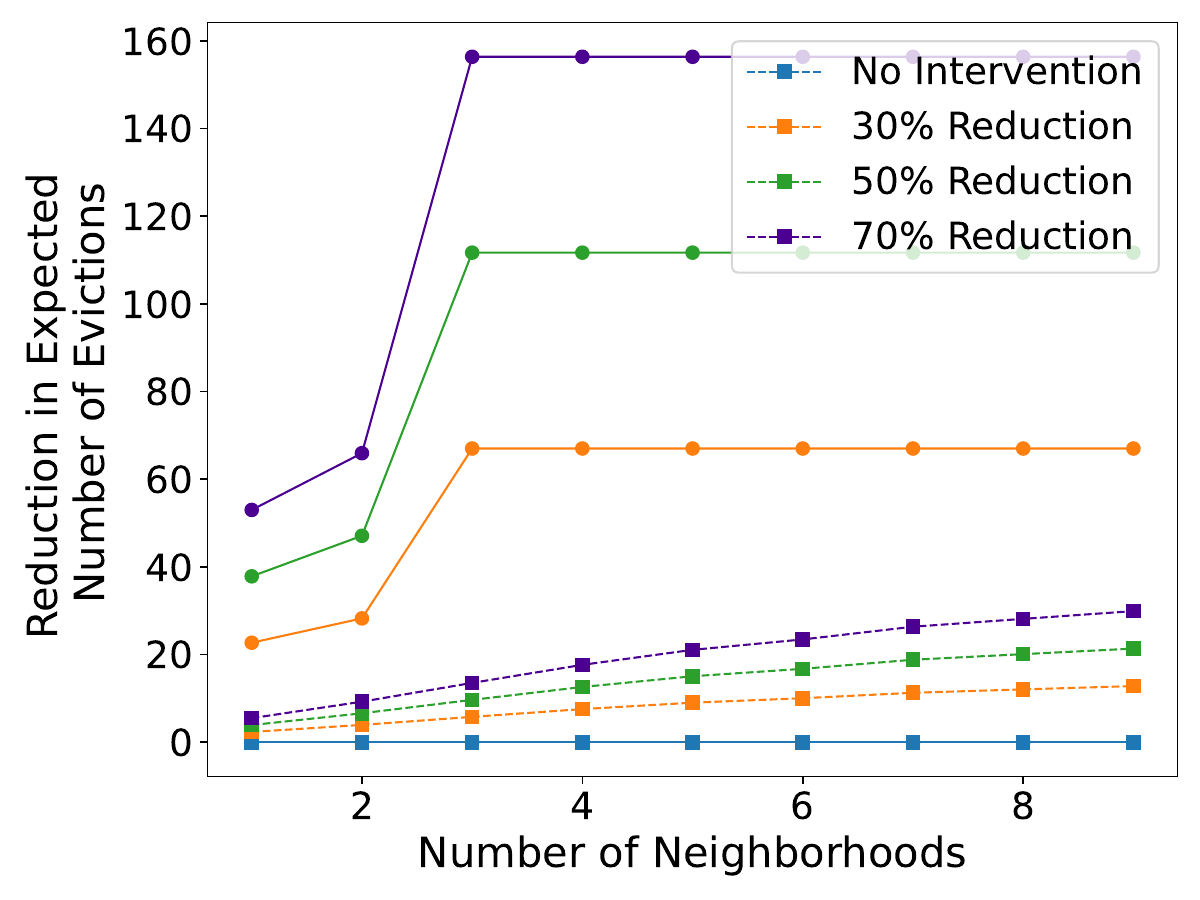}\label{fig:exp_evic_county}}

    \caption{We assess utility — defined as the reduction in the expected number of evictions — across various levels of reduction in eviction probability and numbers of canvassed neighborhoods. In the plots, the X-axis shows the number of neighborhoods canvassed for the non-targeting policy (or equivalent cost travel for the targeting policy), while the Y-axis indicates the reduction in expected evictions for targeting (solid lines) and non-targeting (dashed lines) policies. In both St. Louis City~(a) and County~(b), the targeting approach consistently yielded greater reductions in evictions.
    \label{fig:comparison_calibration}}
\end{figure}

As the allocated total cost increases, canvassers can visit more properties. Both types of policies lead to better outcomes with higher reductions in the expected number of evictions. However, targeting policies (solid lines) result in larger reductions in expected evictions compared with non-targeting policies (dashed lines) (Figure~\ref{fig:comparison_calibration}).

For example, when the probability of eviction is reduced by $30\%$ and the budget is set to the cost of visiting all properties in up to $3$ neighborhoods, the targeting policy leads to an expected reduction of $67$ evictions in St. Louis County, while the non-targeting policy only leads to an expected decrease of $12$ evictions (Figure~\ref{fig:exp_evic_county}). As the figure shows, the benefits of targeting become even more noticeable as the reduction in the probability of eviction increases.
We also observe similar results with the data on St.~Louis City~(Figure~\ref{fig:exp_evic_city}).

\section{Discussion} This work investigates resource allocation in the context of spatial inequalities of risk in geographic units. We propose a general framework based on the classic Mallows model to replicate such inequalities in simulation. To control the degree of inequality, we utilize the dispersion parameter of the Mallows model. We analyze and compare both prediction-based targeting and non-targeting allocation policies in the presence and absence of inequality. Our in-depth simulation settings and experiments show that the targeting policy outperforms the non-targeting policy across different resource allocation scenarios. The conditions include, but are not limited to, low-capacity, high-capacity, low-cost, high-cost, etc. 

Finally, we apply our general framework to St. Louis, MO, known for its racial and socioeconomic segregation. We show that using our framework can reasonably replicate the distribution of risk within St. Louis City and County. Thus, the comparative results demonstrated in simulated settings hold for St. Louis by association. %Furthermore, we show how prediction risk scores can be used for targeting in the context of observing and reducing evictions. 

Although we present significant findings on spatial inequality and its impact on allocation policies, the results must consider key limitations. Our framework is based on the classic Mallows model of ranking; while the ability to calibrate these models to at least two real, albeit neighboring, urban areas (St. Louis City and County) is promising, it will be important to understand what happens with alternative ranking models as well. Testing the framework for other urban areas will be an interesting extension of the work. Additionally, our framework models inequality in terms of eviction risk; yet, inequality can be defined based on many other features, such as welfare, aggregated income, median rent, etc. Replicating other societal resource allocation settings with limited resources that incorporate a risk-score-based preference metric using our framework represents an exciting application and provides opportunities for further testing in the future. We are confident that policymakers can evaluate both targeting and non-targeting context-aware policies using our framework. 

\section*{Ethical Statement}
Our study is based on an initial model designed to compare allocation policies to help policymakers evaluate AI-powered targeting and non-targeting policies, in the context of social service provision. As a case study, we demonstrate the application of our model to tenant eviction prevention through the allocation of outreach case workers. We are not specifically advocating for the use of any particular policy in practice. The main objective of our work is to assess and evaluate policies in terms of how well they perform in assisting at-risk tenants in the presence of inequality. In the real-world context, the specific features of different approaches (for example, leaving out certain neighborhoods to reach more individuals) would need to be carefully considered by stakeholders.

\section*{Acknowledgments}
We are grateful for support from NSF Award 2533162. We also acknowledge the St. Louis Regional Response Team and members of the collaborative network who offer ongoing assistance in the research design,
including our civic technology partner who assembled the administrative records.
\bibliography{references}
\end{document}